\documentclass[aip,jcp,twocolumn,reprint]{revtex4-1}
\usepackage{graphicx}%
\usepackage{dcolumn}%
\usepackage{bm}%
\usepackage{color}
\usepackage{multirow}
\usepackage[normalem]{ulem}
\usepackage{amsmath}
\usepackage{enumerate}
\usepackage{amsfonts}
\usepackage{epsfig}
\usepackage{graphicx}
\usepackage{cancel}
\usepackage[caption=false]{subfig}
\usepackage{floatrow}
\usepackage{subfig}
\newcommand{\be}{\begin{equation}}
\newcommand{\ee}{\end{equation}}
\newcommand{\bea}{\begin{eqnarray}}
\newcommand{\eea}{\end{eqnarray}}

\definecolor{green}{rgb}{0.0, 0.44, 0.0}
\definecolor{red}{rgb}{1.0, 0.13, 0.32}
\definecolor{blue}{rgb}{0.06, 0.2, 0.65}
\definecolor{darkgreen}{rgb}{0,0.5,0}
\definecolor{darkblue}{rgb}{0,0,0.6}
\definecolor{purple}{rgb}{0.4,.2,0.7}
\definecolor{magenta}{rgb}{1.0,0.0,1.0}
\usepackage[colorlinks=true,citecolor=darkgreen,linkcolor=purple,urlcolor=purple]{hyperref}

\def\le{\left}
\def\ri{\right}

\def\qc{q_{\rm c}}

\begin{document}
\floatsetup[figure]{style=plain,subcapbesideposition=top}

\title{Efficient measurement of point-to-set correlations and overlap fluctuations in glass-forming liquids} 

\author{Ludovic Berthier}
\affiliation{Laboratoire Charles Coulomb, UMR 5221 CNRS and 
Universit{\'e} de Montpellier, Montpellier, France}

\author{Patrick Charbonneau}
\affiliation{Department of Chemistry, Duke University, Durham, 
North Carolina 27708, USA}

\affiliation{Department of Physics, Duke University, Durham, 
North Carolina 27708, USA}

\author{Sho Yaida}
\email{sho.yaida@duke.edu}
\affiliation{Department of Chemistry, Duke University, Durham, 
North Carolina 27708, USA}

\begin{abstract}
Cavity point-to-set correlations are real-space tools to detect the roughening of the free-energy landscape that accompanies the dynamical slowdown of glass-forming liquids.  Measuring these correlations in model glass formers remains, however, a major computational challenge. Here, we develop a general parallel-tempering method that provides orders-of-magnitude improvement for sampling and equilibrating configurations within cavities. We apply this improved scheme to the canonical Kob-Andersen binary Lennard-Jones model for temperatures down to the mode-coupling theory crossover. Most significant improvements are noted for small cavities, which have thus far been the most difficult to study. This methodological advance also enables us to study a broader range of physical observables associated with thermodynamic fluctuations.  We measure the probability distribution of overlap fluctuations in cavities, which displays a non-trivial temperature evolution. The corresponding overlap susceptibility is found to provide a robust quantitative estimate of the point-to-set length scale requiring no fitting. By resolving spatial fluctuations of the overlap in the cavity, we also obtain quantitative information about the geometry of overlap fluctuations. We can thus examine in detail how the penetration length as well as its fluctuations evolve with temperature and cavity size.
\end{abstract}

\maketitle
\section{Introduction}
A well-known difficulty in understanding the increasing sluggishness of glass-forming liquids upon lowering temperature $T$ is that no obvious change to the static structure of the liquid accompanies it~\cite{BB11}.
In stark contrast to the critical slowing down observed near a standard critical point, static correlation functions of glass formers barely budge while the structural relaxation timescale grows by more than $15$ orders of magnitude.
The glassy slowdown has thus instead been attributed to a different type of criticality, one at which the free-energy landscape becomes very rugged, leading to the emergence of many metastable states separated 
by growing free-energy barriers~\cite{KTW89,MP99,LW07}.

In order to capture this ruggedness, the concept of a point-to-set (PTS) correlation was introduced about a decade ago~\cite{BB04,MS06}.
PTS correlations generalize multi-point correlations to their infinite-point limit, which makes them sensitive to the structure of the full configuration.
Various PTS geometries have since been considered, including 
cavity~\cite{CGV07,BBCGV08,SL11,HMR12,BKP13}, 
random pinning~\cite{CCT12,CB12,JB12,CCT13,CT13,KB13,OKIM15}, and walls~\cite{KVB12,HBKR14}.
Despite subtle differences between them~\cite{BK12,BICtest12}, the overarching observation remains.
For glass formers the PTS correlation length grows more rapidly than lengths extracted from two-point correlation 
functions~\cite{BBCGV08,BK12,CCT12,HMR12,HBKR14} and is thus an important aspect of glassy physics. However, extracting PTS correlations from numerical simulations has thus far been limited by the computational challenge of measuring them~\cite{BICtest12,BK12}.

In order to better understand the nature of this challenge, let us consider the case of PTS correlation within the cavity geometry.
First consider a liquid at equilibrium, and specify a finite cavity, say a sphere of radius $R$, and pin everything outside of it -- thus fixing a \emph{set} of particles.
Then allow the particles inside the cavity to explore phase space under the constraints exerted by the set of pinned particles, now acting as an effective quenched disorder.
Particles inside the cavity thus explore the local free-energy landscape. A standard way of characterizing these local landscapes, adapted from the study of spin glasses, is to consider the statistics of the overlap, $q\le({\bf r}\ri)$, between two equilibrium configurations at a \emph{point} ${\bf r}$ inside the cavity.
The materials we consider are characterized by a very slow dynamics in the bulk, and confining them with amorphous boundaries is found to make the natural dynamics of these systems even slower~\cite{BICtest12,BK12,HBKR14,CGB13}. 
The major computational problem is thus that of properly sampling the equilibrium fluctuations of the confined fluid.

The origin of the dynamical slowdown may be attributed to the confining amorphous boundaries provoking an entropy crisis qualitatively similar to the one which could be happening at the Kauzmann temperature, $T_{\rm K}$, in the bulk liquid~\cite{K48}. 
Increasing the confinement would reduce the number of accessible states from being exponentially large in the number of particles for large cavities, to being sub-exponential for small cavities~\cite{BB04}.
This shift would then be evidenced in the evolution of overlap statistics in the local landscapes, which is exactly what PTS correlations purport to capture, and its location defines the PTS correlation length, $\xi_{\rm PTS}(T)$.
From this perspective, the computational difficulty encountered in previous studies for $R \lesssim \xi_{\rm PTS}$ should not come as a surprise.
Confinement should be inducing the analog of an equilibrium (or `ideal') glass transition in a system containing a finite number of particles~\cite{BB04}.
Proper sampling then requires very large computer resources, as recent studies of constrained systems have evidenced~\cite{KB13,OKIM15,BC14}.
As a result the entropy crisis picture has only been partially validated.
In particular, although solid evidence for a growing PTS correlation length has been obtained in various systems~\cite{BBCGV08,SL11,HMR12,CCT12,BKP13}, several questions remain unexplored regarding the nature of the crossover between high and low overlap regimes, the associated fluctuations, and the connection to dynamical relaxation in the bulk.

In this paper we develop a generic computational method to more quickly and reliably sample PTS correlations.
As a proof of principle, we study the Kob-Andersen binary Lennard-Jones model~\cite{KA94,KA95}, which is a classical glass-forming liquid.
Yet our approach is sufficiently generic to be applied to other glassy systems, including hard spheres. 
We use the efficiency of our approach to record novel physical observables, beyond the PTS correlation function, that rely on efficiently sampling thermodynamic fluctuations inside a cavity.
In particular, we follow the evolution with cavity size and temperature of the complete probability distribution function of the overlap.
Its variance defines an overlap susceptibility that allows us to directly locate the PTS crossover scale without any empirical fitting.
In addition, we are able to spatially resolve overlap fluctuations inside the cavity, giving us access both to their radial profile and their orthoradial fluctuations, and thus to make connection with PTS measurements in other geometries, such as flat walls.

The plan of this paper is as follows. In Sections~\ref{sec:model} and~\ref{sec:PT} we detail the simulation model and the parallel-tempering methodology, respectively. In Section~\ref{sec:PTS_methodology} we present the approach for computing PTS correlations as well as the results, and in Sec.~\ref{sec:penperp} we evaluate results for the wall and wandering lengths from the cavity PTS setup. We conclude in Section~\ref{sec:conclusion}.

\begin{figure*}
\centering{
\sidesubfloat[]{\includegraphics[width=0.45\textwidth]{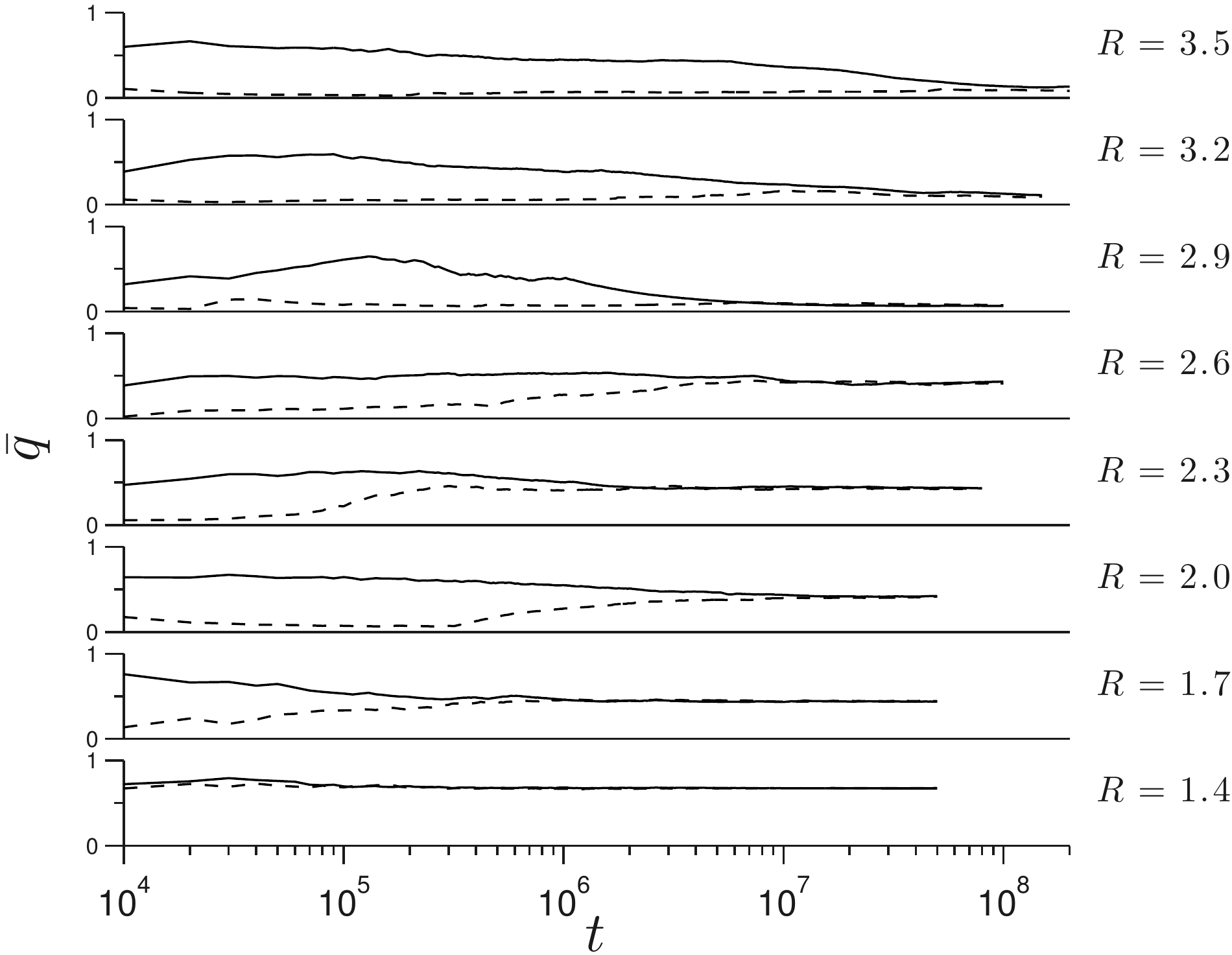}}\quad%
\sidesubfloat[]{\includegraphics[width=0.45\textwidth]{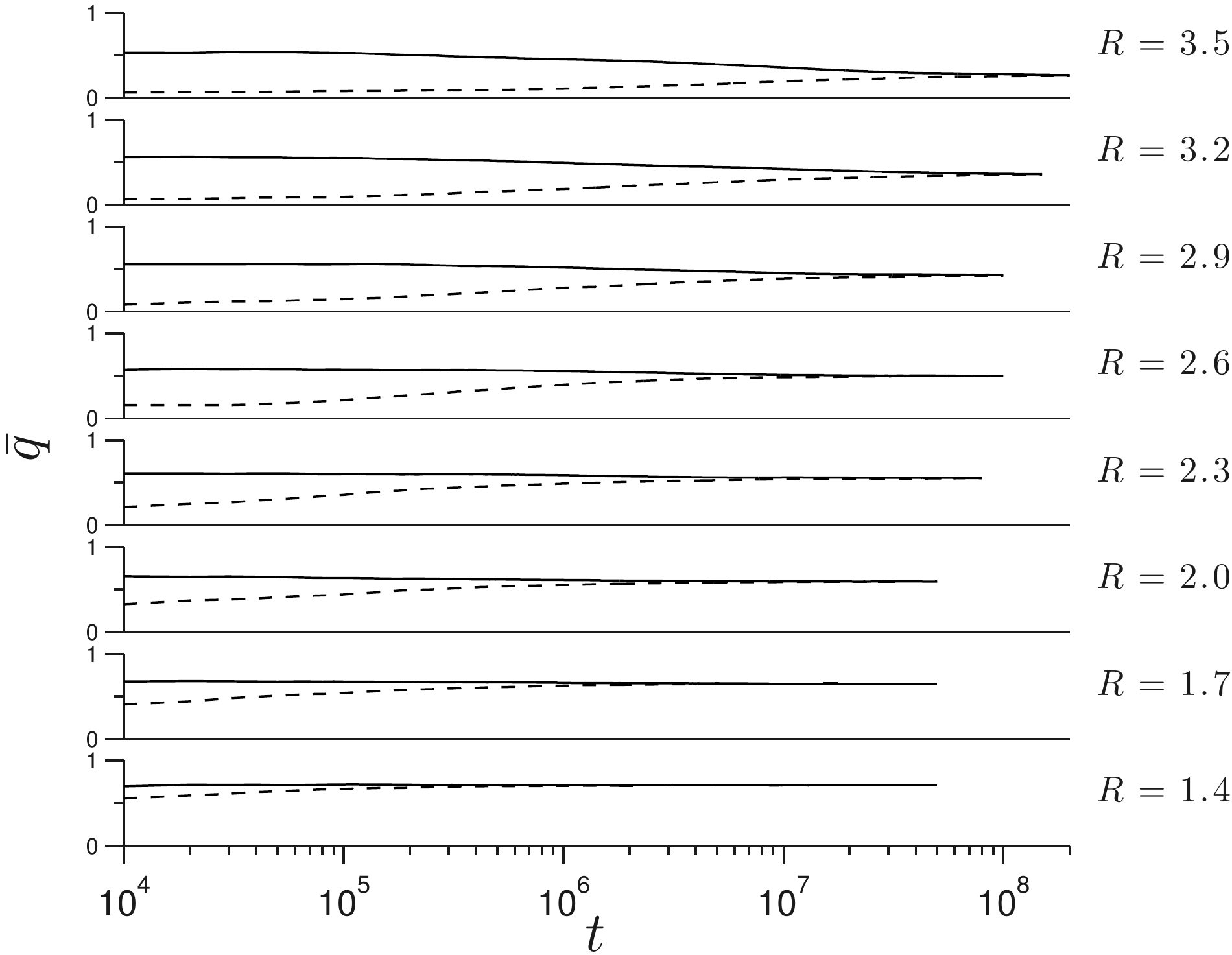}}
}
\caption{(a) Running average of core overlaps, $\bar{q}$, after $t$ MC sweeps from both the original (solid lines) and a randomized (dashed lines) configurations at $T=0.51$ for a cavity of radius $R$. (b) The same quantities averaged over $50$ cavities.
Note that convergence is systematically faster in small cavities, where parallel-tempering 
is more efficient.}\label{BIC}
\end{figure*}

\section{Simulation Model}
\label{sec:model}

We study the behavior of a glass-forming liquid first proposed by Kob and Andersen\cite{KA94,KA95}. 
The Kob-Andersen binary Lennard-Jones (KABLJ) model contains two particle species, denoted $A$ and $B$, with equal mass $m$, and interacting via the pair potential
\be\label{LJ}
V_{\alpha\beta} (r) = 4 \varepsilon_{\alpha\beta} \le[ \le(\frac{\sigma_{\alpha\beta}}{r}\ri)^{12} - \le(\frac{\sigma_{\alpha\beta}}{r}\ri)^6 \ri]\, ,
\ee
where, $\alpha,\beta \in \{A, B\}$ with parameters $\varepsilon_{AB} / \varepsilon_{AA}= 1.5$, $\varepsilon_{BB} / \varepsilon_{AA} = 0.5$, $\sigma_{AB} / \sigma_{AA} = 0.8$, and $\sigma_{BB}/\sigma_{AA} =0.88$.
The interaction potential is cut off at $r^{\rm cut}_{\alpha\beta}=2.5 \sigma_{\alpha\beta}$ and shifted, such that it vanishes at the cutoff.
The relative number of particles is $N_A:N_B = 4:1$ and the overall number density is $\rho=N/V=1.2\sigma_{AA}^{-3}$.
Note that we report below lengths and temperatures in standard dimensionless Lennard-Jones units set by $\sigma_{AA}$ and $\varepsilon_{AA}/k_{\rm B}$.

We use bulk samples with $N=135,000$ particles in a periodic cubic box, generated as described in Ref.~\onlinecite{DLY13}, at temperatures $T=1.00, 0.80, 0.60, 0.51, 0.45$.
At each temperature, we take $50$ equilibrated snapshots, separated by more than $2 \tau_{\alpha}$, where $\tau_{\alpha}$ is the bulk structural relaxation time.
Then, within each snapshot, we randomly select a position as cavity center.
For each center, we fix the particles outside the cavities of radius $R$, and continue sampling configurations for the $N_{\rm cav}$ particles inside the cavity, as described in Sect.~\ref{sec:PT}.

Note that in order to directly evaluate how the behavior of the cavity changes as function of $R$, we preserve each of the cavity centers as $R$ increases. Note also that the system size used here is far larger than necessary, because the length scales we extract in Sections~\ref{sec:PTS_methodology} and~\ref{sec:penperp} are rather small. Sampling $50$ cavity centers within a single snapshot would have sufficed to ensure statistical independence of the resulting cavities, and would have thus provided similar results.

\section{Parallel-tempering methodology and convergence}
\label{sec:PT}

A key hurdle to measuring PTS observables in simulations has been the large computational effort needed to sample different equilibrium configurations inside a cavity.
Here, we design a Monte Carlo parallel-tempering~\cite{FS01} scheme (a type of Hamiltonian replica exchange\cite{FWT02}) that sidesteps this difficulty by dramatically lowering barriers within the rugged free-energy landscape. The approach goes as follows.
We prepare $a=1,...,n$ replicas of a given configuration inside the cavity. The replicas evolve at different temperatures $T_{a}$ and `shrinkage' parameters $\lambda_{a}\leq1$ (see the Appendix for actual values) with a deformed Hamiltonian
\bea
V_{\alpha\beta} (r; \tilde{\lambda}_a) = 4 \varepsilon_{\alpha\beta} \le[ \le(\frac{\tilde{\lambda}_a\sigma_{\alpha\beta}}{r}\ri)^{12} - \le(\frac{\tilde{\lambda}_a\sigma_{\alpha\beta}}{r}\ri)^6 \ri]\, ,
\eea
where $\tilde{\lambda}_a=\lambda_{a}$ for a pair of mobile particles within a cavity and $\tilde{\lambda}_a=\frac{1+\lambda_{a}}{2}$ for a pair containing one mobile (inside the cavity) and one pinned particle (outside the cavity).
We draw cavity configurations from the bottom replica with $T_1=T$ and $\lambda_1=1$, whereas the other replicas evolve at higher temperatures $T_a  > T$ and 
with smaller particles, $\lambda_{a}<1$, in order to speed up their 
dynamics. 

 Within each replica, we perform simple Monte Carlo (MC) moves that consist of: (i) choosing a particle $i$ from $N_{\rm cav}$ mobile particles inside the cavity; (ii) displacing particle $i$ by $\Delta {\bf x}=l \hat{\bf{n}}$, where $l$ is uniformly drawn from $[0,0.3]$ and $\hat{\bf{n}}$ uniformly drawn on the sphere $S^2$;  and (iii) accepting/rejecting the displacement according to the Metropolis criterion. Note that we put a hard spherical wall at the edge of the cavity, so that all moves that take a mobile particle outside the cavity radius are rejected. Each MC sweep consists of $N_{\rm cav}$ MC trial moves, hence on average each particle attempts to move once.
More crucially, in order to release the disorder constraint (frustration) induced by the pinned particles, identity-exchange of a pair of adjacent replicas [in $(T_a,\lambda_a)$ space] is attempted every $1000$ MC sweeps, on average.
This replica swapping is again accepted or rejected according to the Metropolis criterion, so that our MC algorithm ensures proper equilibrium sampling.
Compared to earlier schemes for cavity studies, the proposed method 
is distinct from the simpler annealing procedure used before for the 
same model~\cite{HMR12}, and is 
more generally applicable than the local particle swap Monte Carlo moves that 
are only efficient for specific glass-forming models~\cite{BBCGV08}.

The quality of the equilibration within each cavity is evaluated by employing two schemes, following the approach developed by 
Cavagna {\it et al.}~\cite{BICtest12}: we start the system from (i) the original configuration and (ii) a randomized configuration prepared by putting the cavity at $T=1.00$ and $\lambda=0.6$ for $10^4$ MC sweeps. We then record the re-equilibrated configurations every $t_{\rm rec}=10^4$ MC sweeps and monitor the core overlap (as defined in Sect.~\ref{sec:PTS_methodology}) between the new configuration and the original cavity configuration, $\qc^{\rm on}\le(t\ri)$, as a function of the number of MC sweeps, $t$. Their running average
\be
\bar{q}(t)\equiv\frac{1}{\le(t/t_{\rm rec}\ri)}\sum_{s=1}^{\le(t/t_{\rm rec}\ri)}\qc^{\rm on}\le(t_{\rm rec}s\ri)\, 
\ee
decreases for the first scheme and increases for the second, converging upon equilibration (Fig.~\ref{BIC}).
The first $s_{\rm eq}$ configurations are discarded, and the overlap for the following $s_{\rm prod}$, 
\be\label{qon}
\langle \qc^{\rm on}\rangle\equiv\frac{1}{s_{\rm prod}}\sum_{s=s_{\rm eq}+1}^{s_{\rm eq}+s_{\rm prod}}\qc^{\rm on}\le(t_{\rm rec}s\ri)
\ee
is computed. Convergence is deemed obtained when the results of both approaches lie within $\pm q_{\rm tol}$ of each other for each cavity. Replica parameters as well as $s_{\rm eq}$ and $s_{\rm prod}$ are chosen, such that for $q_{\rm tol}=0.1$, at least $98\%$ of cavities pass this convergence test (see Appendix for actual values). However, because the difference between the two approaches is not systematic, averaging over $50$ cavities results in a very close agreement between the two schemes, namely in a convergence of $\bar{q}(t)$ within $\pm0.005$.

This algorithm behaves differently in different cavity size regimes.
The most impressive results are obtained for small cavities, even at low temperatures. Only a few replicas are indeed needed for a small system to jump over the relatively high free-energy barriers and thus quickly equilibrate and properly sample configurations. This efficiency is illustrated in Fig.~\ref{BIC}, which shows that the algorithm time to convergence is smaller for smaller cavities.
This outcome markedly contrasts with simpler MC schemes, the sampling efficiency of which is akin to that of molecular dynamics.
As an example, results from Ref.~\onlinecite{HMR12} suggest that for low temperatures $T\leq0.51$ and small radii $R\leq 3.0$,  at least $10^{10}$ MC sweeps would be needed to equilibrate the system. Here, by contrast, we can equilibrate a similar system within $\sim 10^7$ MC sweeps using $\sim 10$ replicas.
(Note, however, that the computational efficiency of our algorithm depends sensitively on the details of the parallel-tempering parameters, which require more fine-tuning than the parameters of simpler schemes.)
We thus conservatively achieve at least a $100$-fold speedup, and that speedup would be even stronger for state points where simpler MC schemes have not been found to converge. 
By contrast, for $R\gg\xi_{\mathrm{PTS}}$, confinement effects are negligible and even simple Monte Carlo moves suffice to sample configurations.
The most problematic regime is $R\sim\xi_{\mathrm{PTS}}$ at low temperatures, where even relatively large cavities display fine-tuning bottlenecks.
That this regime sets the lower limit on the convergence criterion is unsurprising, as the optimal spacing between adjacent replica parameters generically scales as $1/\sqrt{N_{\rm cav}}$. We thus need $\sim\xi_{\mathrm{PTS}}^{3/2}$ replicas to properly equilibrate a sample at this most resilient regime.

In earlier work~\cite{BICtest12}, 
an acceleration of the overlap dynamics was reported 
in a binary soft sphere mixture, where traditional Monte Carlo 
moves are supplemented by binary particle exchanges (or `swap' moves),
a method which was used in a series of numerical 
works~\cite{CGV07,BBCGV08,BICtest12}. 
We cannot directly compare 
our work to these dynamic measurements, as our parallel-tempering scheme 
does not allow us to measure time correlation functions at 
fixed temperature,
and no evolution of the convergence time was reported using particle 
swaps only~\cite{BICtest12}. We believe, 
however, that the speedup of the convergence time that we report 
in Fig.~\ref{BIC} is of a different nature, and should uniquely  
be attributed to the merits of the parallel-tempering scheme
and not to any natural dynamical process taking 
place in the glass-forming liquid at various degrees of confinement.
Methods developed in Ref.~\onlinecite{victor1} could be used in the future to 
quantify more precisely the convergence time of the 
parallel-tempering scheme using time-correlation functions associated to 
the random walk of the various copies in the $(T_a,\lambda_a)$ space. 

\section{Point-to-set observables}
\label{sec:PTS_methodology}

\begin{figure*}
\centering
\sidesubfloat[]{\includegraphics[width=0.27\textwidth]{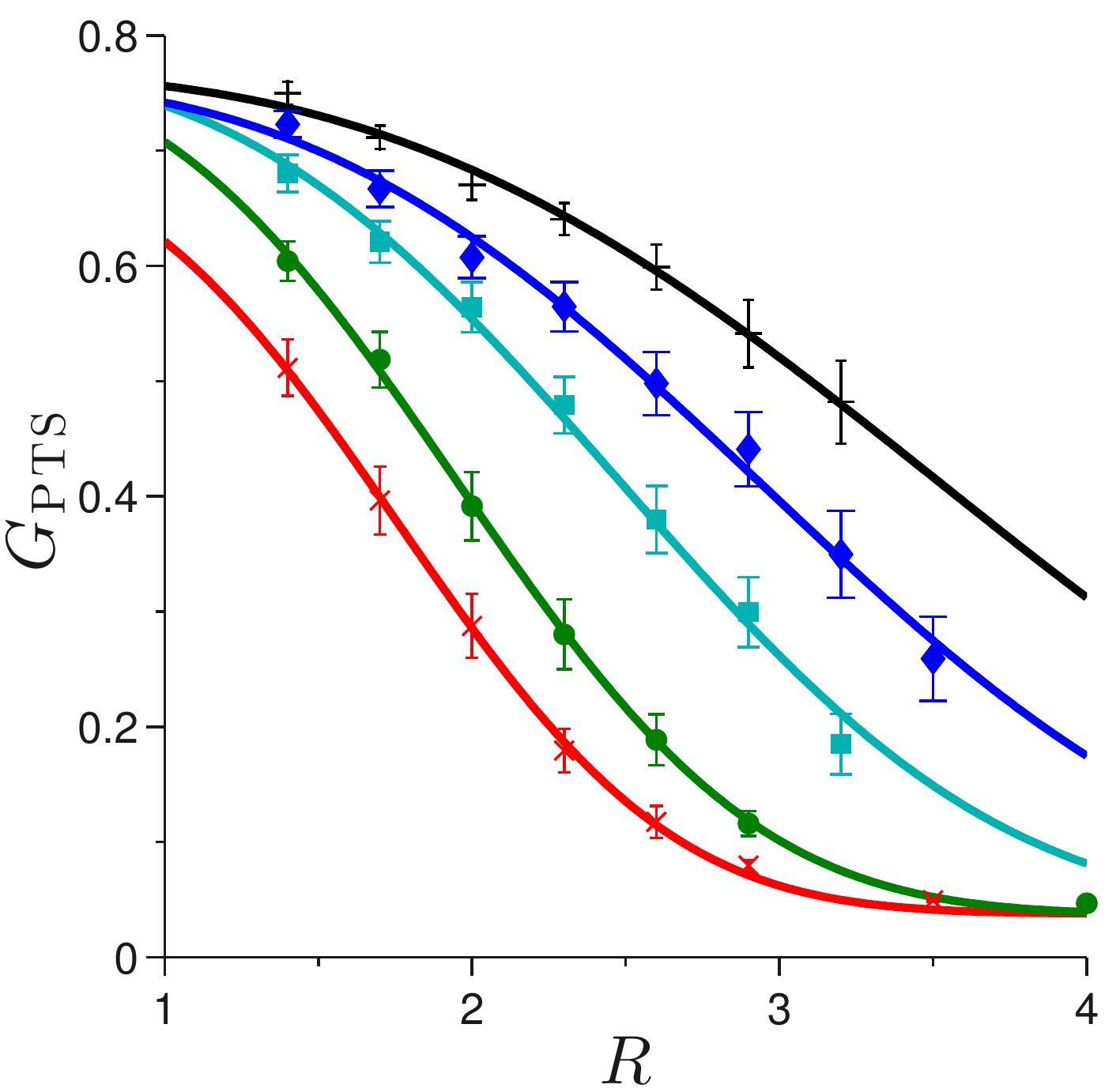}\label{PTS}}\quad%
\sidesubfloat[]{\includegraphics[width=0.27\textwidth]{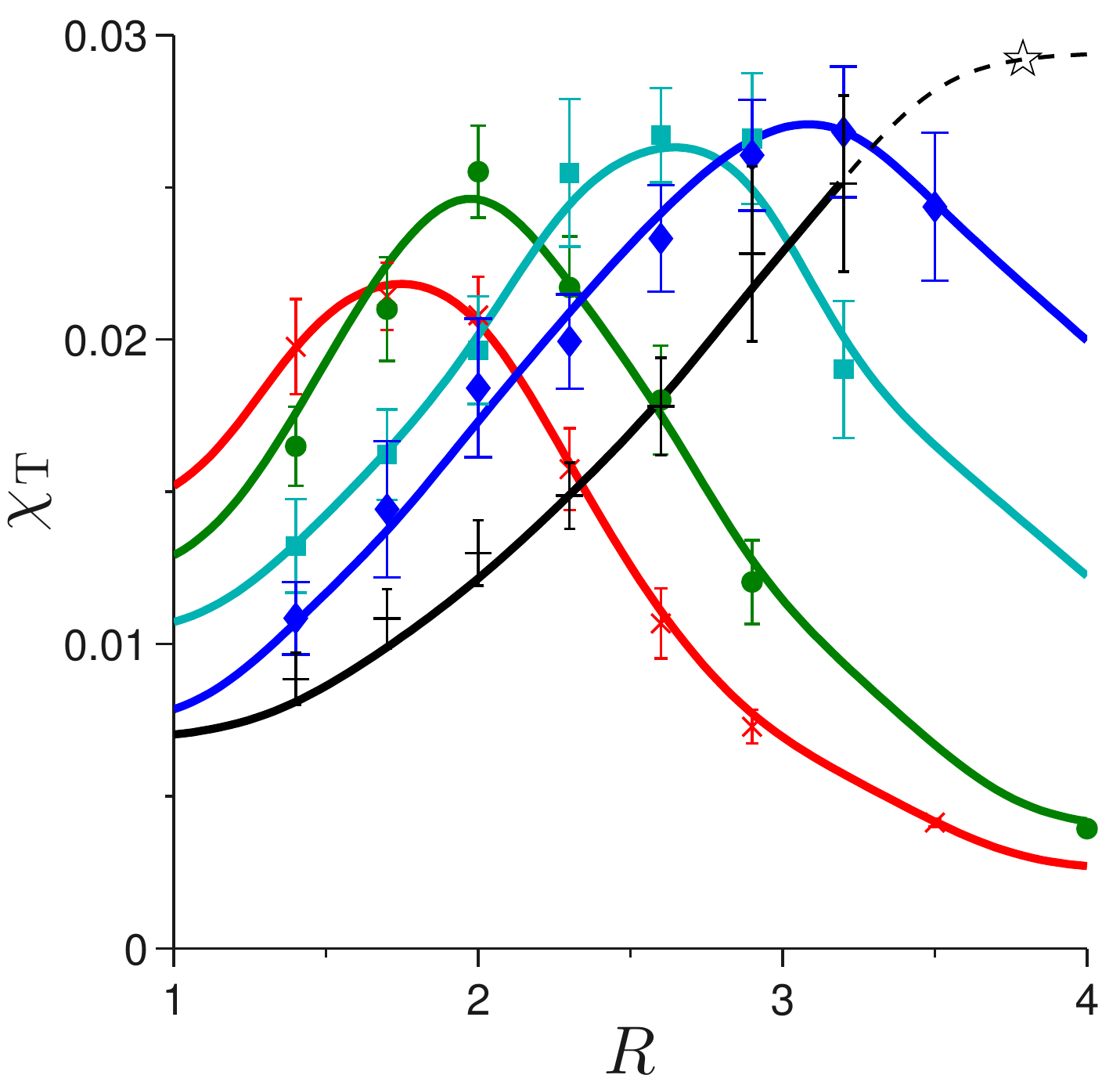}\label{ChiT}}\quad%
\sidesubfloat[]{\includegraphics[width=0.27\textwidth]{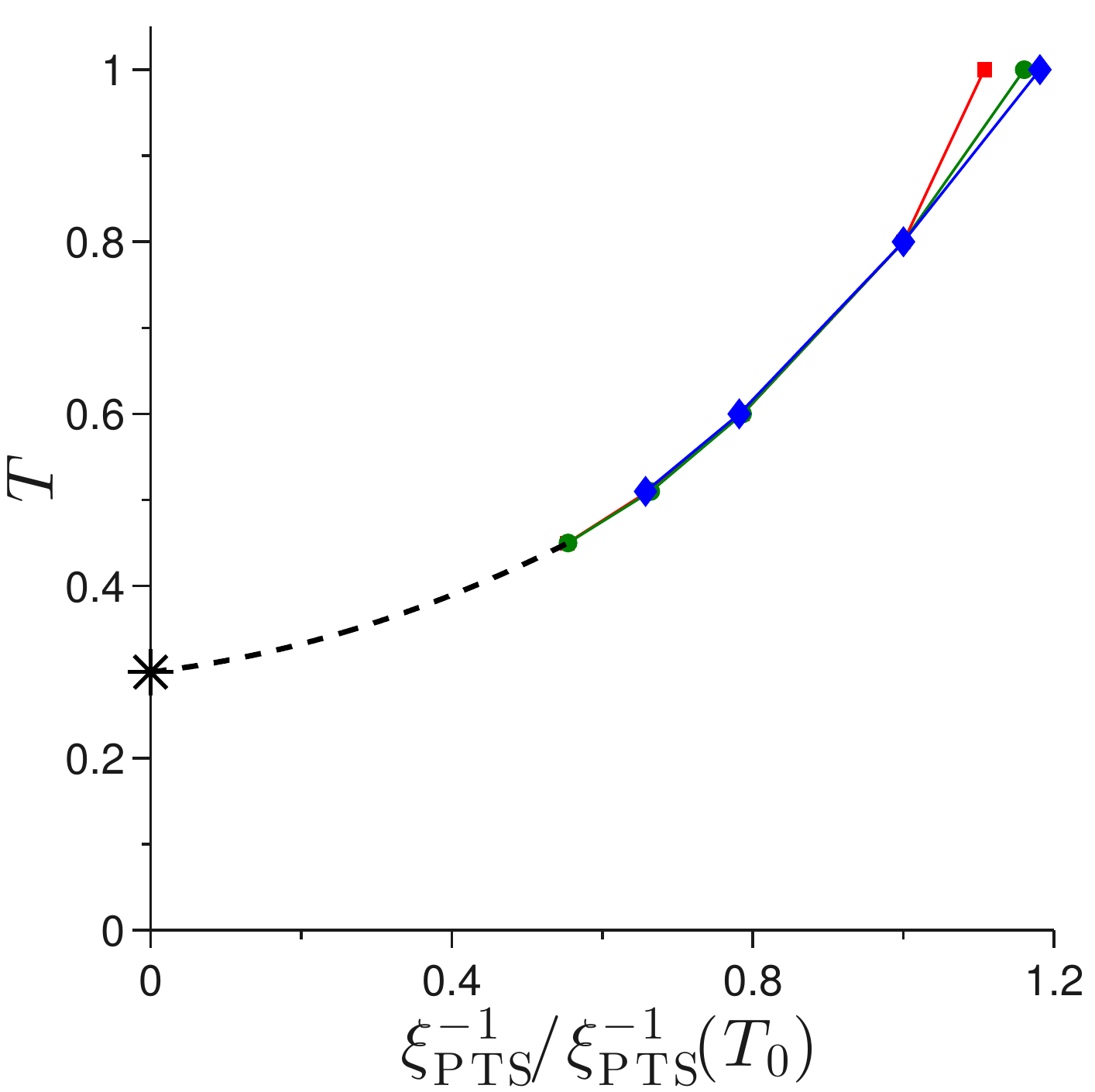}\label{TK}}%
\caption{
(a) Radial decay of the cavity PTS correlation at $T = 1.00$ (red-cross), $0.80$ (green-circle), $0.60$ (cyan-square), $0.51$ (blue-diamond), and $0.45$ (black-plus).
Solid lines are fits to a compressed exponential, $G_{\mathrm{PTS}}=A\exp[-(R/\xi_{\rm PTS}^{\rm fit})^{\eta}]+G_{\rm PTS}^{\rm bulk}$, where the bulk overlap value, $G_{\rm PTS}^{\rm bulk}$, is obtained from $4000$ pairs of independent configurations in bulk samples, and $\eta=3$ is found to work reasonably well for all $T$.
(b) PTS susceptibilities with cavity radius $R$. Solid lines are guides for the eyes. The dashed line for $T=0.45$ is extrapolated from $R=3.2$ to the extrapolated peak (star) at $R\approx3.8$. See text for details.
(c) Temperature evolution of inverse $\xi_{\rm PTS}^{\rm fit}$ (red-square), $\xi_{\rm PTS}^{\rm th}$ (green-circle), and $\xi_{\rm PTS}^{\rm peak}$ (blue-diamond), rescaled to unity at $T_0=0.8$, along with an extrapolation to prior estimates of bulk $T_\mathrm{K}\approx0.30$ for this model (dashed line)~\cite{OKIM15,SKT99}.
This representation emphasizes the similarity with other constrained phase diagrams, such as random pinning and coupling to a reference configuration.
}
\label{PTSs}
\end{figure*}

Once the glass-forming liquid is confined by amorphous boundaries in 
a finite cavity we need to analyse the equilibrium fluctuations 
of the overlap field $q({\bf r})$  inside the cavity, which  
is the principal observable used in this paper.
To this end,
we proceed as follows. 
Denote a pair of configurations by ${\bf X}=\le\{{\bf x}_i\ri\}$ and ${\bf Y}=\le\{{\bf y}_i\ri\}$.
For each particle ${\bf x}_i$, find the nearest particle ${\bf y}_{i_{\rm nn}}$ of the same species, and assign an overlap value $q_{{\bf X; Y}} \le({\bf x}_i\ri)\equiv w\le(\big|{\bf x}_i-{\bf y}_{i_{\rm nn}}\big|\ri)$, where
\be
w(z)\equiv {\rm exp}\le[-\le(\frac{z}{b}\ri)^2\ri],
\ee
with $b=0.2$.
This function defines overlap values $q_{\bf X; Y} \le({\bf x}_i\ri)$ at scattered points $\le\{{\bf x}_i\ri\}$.
We then define $q_{\bf X; Y} \le({\bf r}\ri)$ to be a continuous function precisely passing through these points.
Specifically, we first perform a Delaunay tessellation of space and, to a point ${\bf r}$ within a simplex spanned by four points $\le\{{\bf x}_i\ri\}_{i=i_1,i_2,i_3,i_4}$, we associate a linearly interpolated value
\be
q_{\bf X; Y} \le({\bf r}\ri)=\sum_{i=i_1,i_2,i_3,i_4}c_i q_{\bf X} \le({\bf x}_i\ri),
\ee
where $\le\{c_i\ri\}_{i=i_1,i_2,i_3,i_4}$ satisfies ${\bf r}=\sum_{i=i_1,i_2,i_3,i_4}c_i {\bf x}_i$ with $\sum_{i=i_1,i_2,i_3,i_4}c_i=1$.
Similarly, we obtain $q_{\bf Y; X} \le({\bf r}\ri)$, and define 
\be q_{\bf X,Y}\le({\bf r}\ri)\equiv \frac{1}{2}\le\{q_{\bf X; Y}\le({\bf r}\ri)+q_{\bf Y; X}\le({\bf r}\ri)\ri\}\, ,
\ee
which provides the overlap near the core of the cavity
\begin{equation}
\qc\equiv\frac{3}{4\pi r_{\rm c}^3}\int_{|{\bf r}|<r_{\rm c}}d{\bf r}\ q_{\bf X,Y}\le({\bf r}\ri),
\end{equation}
where $r_{\rm c}=0.5$ and ${\bf r}={\bf 0}$ is the cavity center. This integral is numerically evaluated by Monte Carlo integration with $10^4$ points.
Note that our results are insensitive to both the choice of overlap function $w(z)$ as long as its range is kept of the order of cage size and to distinguishing or not between species. It is, however,  important for the overlap field, $q_{\bf X,Y}\le({\bf r}\ri)$, to be continuous in order to attain good spatial resolutions within a cavity (see Sect.~\ref{sec:penperp}). We thus avoid using a step function.

While we monitored the overlap between the original configuration and a re-equilibrated configuration in checking for a good equilibration, for the rest of the paper, we evaluate the overlap between two statistically independent re-equilibrated configurations.
Overlaps are computed by comparing two configurations obtained from two different schemes, which guarantees their statistical independence.
In the following, we denote $\langle...\rangle_{J(R)}$, the thermal average inside the cavity with the effective quenched disorder $J(R)$ set by a pinned external configuration, evaluated by averaging over $s_{\rm prod}$ pairs of configurations, and $[...]$ the average over disorder, evaluated by averaging over $50$ disorder realizations.

As alluded to in the introduction, we expect a crossover from high to low core overlap around the cavity PTS length scale.
The PTS correlation function,
\be
G_{\rm PTS}\le(R\ri)\equiv\le[\langle \qc \rangle_{J(R)}\ri]\, ,
\ee
indeed decays on a characteristic length scale, $\xi_{\rm PTS}$, that 
clearly grows as temperature decreases, as shown in Fig.~\ref{PTSs}a. 
This behavior is qualitatively consistent with earlier 
work~\cite{BBCGV08,HMR12}, and reveals
a rarefaction of metastable states leading to the growing PTS length $\xi_{\rm PTS}$, as originally envisioned by Biroli and Bouchaud~\cite{BB04}.

The standard way to quantify $\xi_{\rm PTS}$ has thus far been to fit 
the decay of $G_{\rm PTS} \le (R\ri)$ to a compressed exponential
form $G_{\rm PTS} \sim \exp \left[ -(R/\xi)^\eta \right]$, where 
$\eta$ is an adjustable exponent. 
When reliable information about small cavity is available, however, this {\it ad hoc} fitting form is not computationally optimal.  In order to obtain a more robust determination of the PTS length scale, we can instead appeal to another important aspect of the PTS physics: the appearance of an increasing number of accessible configurations as the cavity size increases. 
Existence of these competing states results in configurational fluctuations, in addition to vibrational fluctuations around these configurations.
Hence, the overlap function can take a larger number of possible values, 
and their relative contribution depends on the probability with which alternate configurations can be accommodated by a given cavity. 
In the large cavity limit, most configurations are different from one another and the overlap is very small with very little fluctuations.
In the interesting crossover regime, however, we expect the probability distribution function (PDF) of core overlaps to display a very broad range of fluctuations,
while for small cavities configurations are all very similar, hence the overlap is high and does not fluctuate much.

\begin{figure*}
\centering
\sidesubfloat[]{\includegraphics[width=0.27\textwidth]{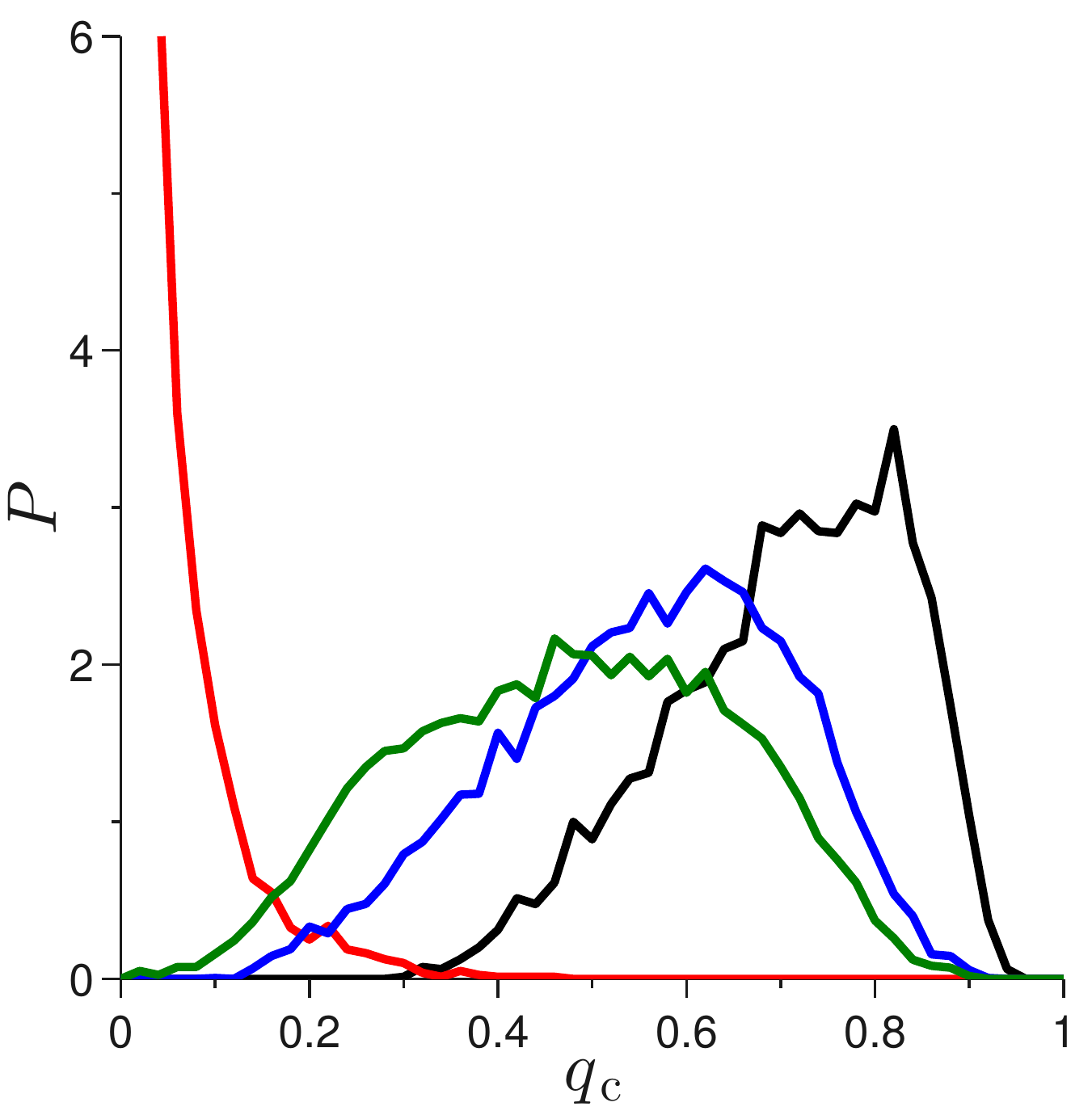}\label{uni}}\quad%
\sidesubfloat[]{\includegraphics[width=0.27\textwidth]{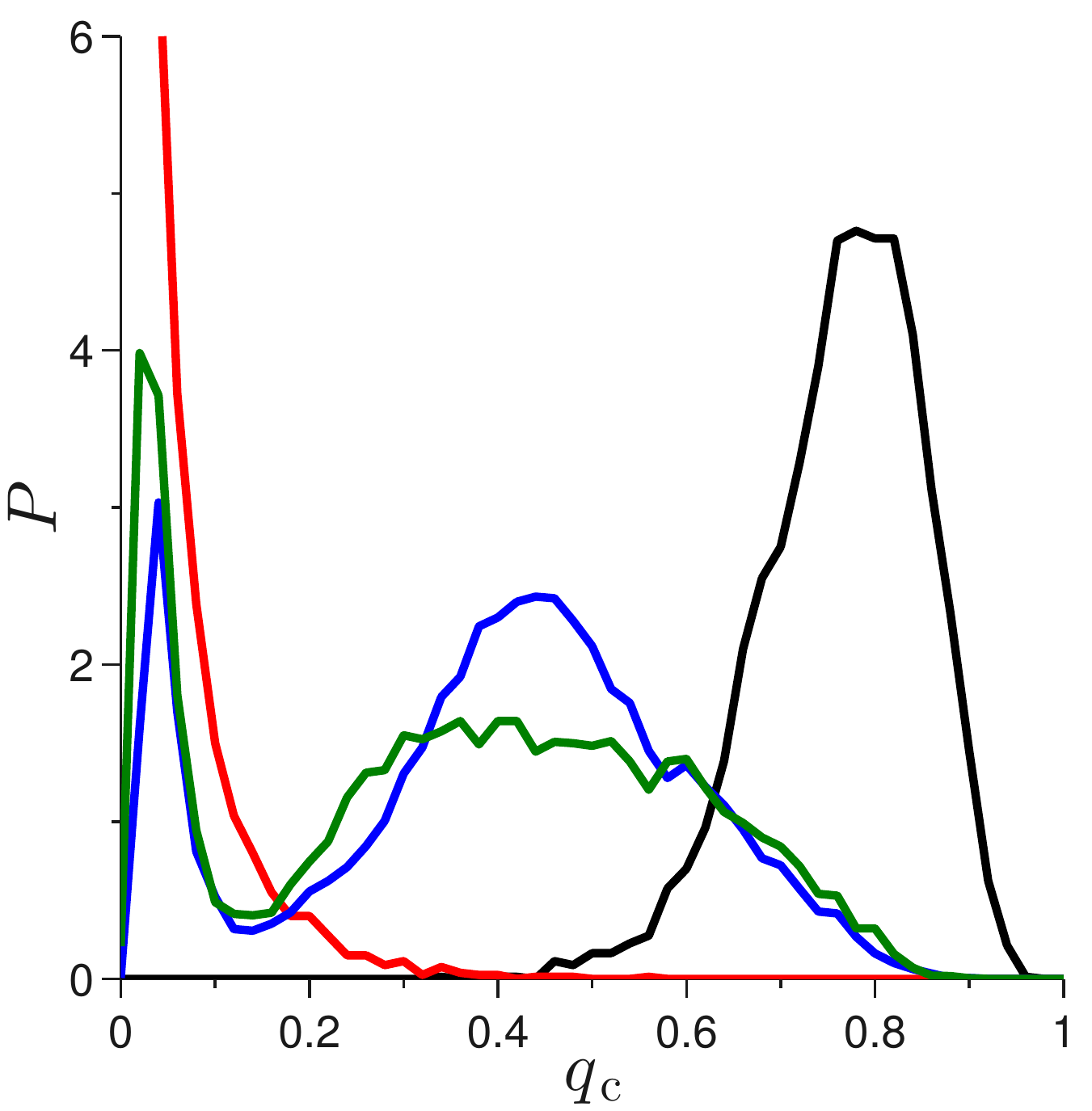}\label{bi}}\quad%
\sidesubfloat[]{\includegraphics[width=0.27\textwidth]{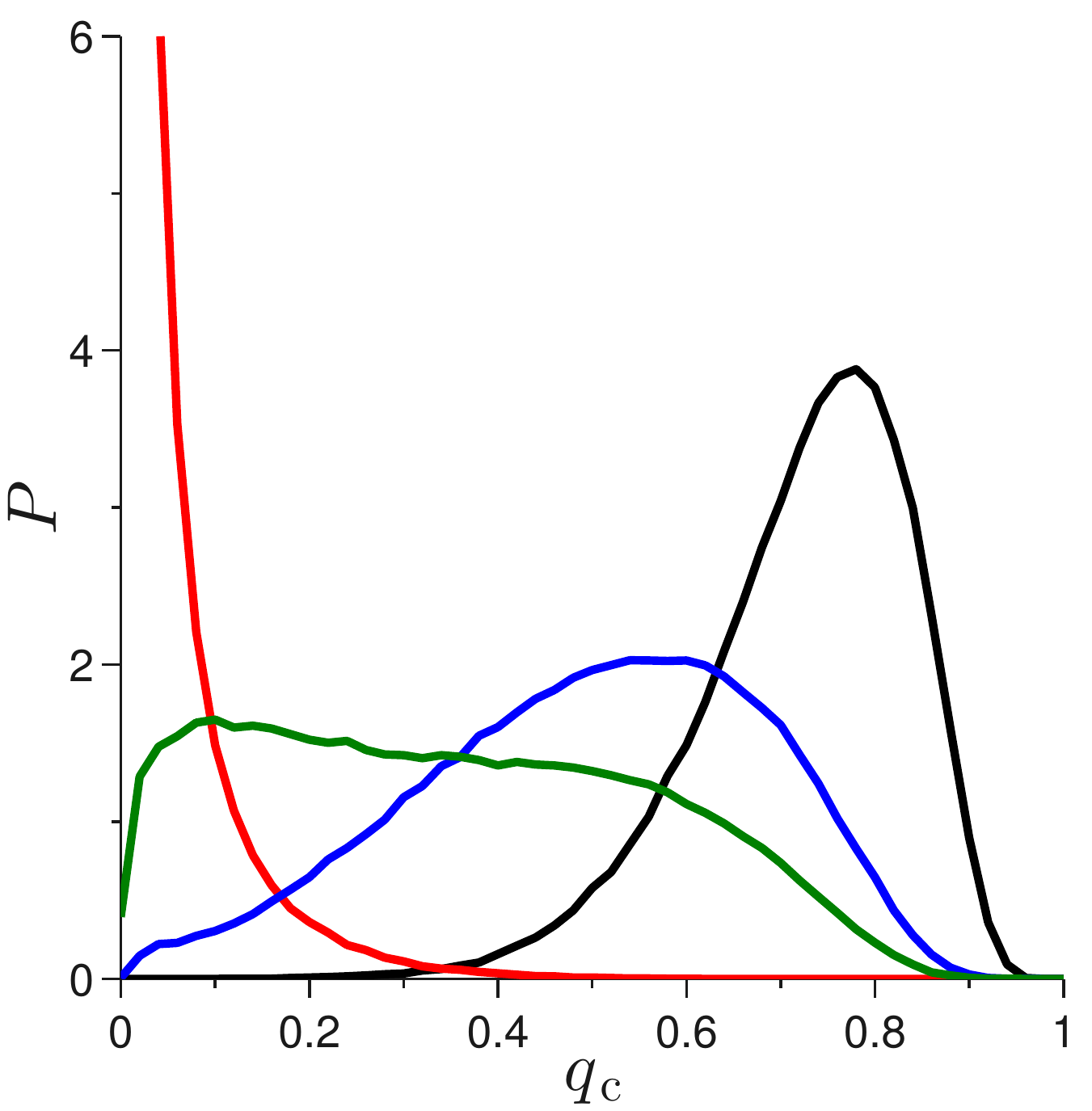}\label{glo}}%
\caption{(a,b) PDF of core overlaps, $P(\qc)$, at $T=0.51$ for $R=1.4$ (black),  $2.6$ (blue), $3.2$ (green), and $6.0$ (red) for two specific cavities. 
As the cavity size increases, the distribution becomes either (a) broad or (b) bimodal around the PTS length scale, depending on the realization of disorder, i.e., the position of the cavity center. (c) The disorder-averaged PDF thus also broadens at the PTS scale and is nearly bimodal at $R \sim \xi_{\rm PTS}$.}
\label{Histogram}
\end{figure*}

These expectations are validated by our numerical results, as illustrated in 
Fig.~\ref{Histogram}. For each cavity and temperature, we find the PDF to be narrow at large and small $R$, with a maximum in the width of the 
fluctuations for an intermediate cavity size. For some (though not all) cavities, 
we find that the PTS crossover size corresponds to a bimodal distribution of overlap values, \textit{e.g.} Fig.~\ref{Histogram}b.
As temperature is lowered, we find that the fraction of bimodal distributions increases, which results in a disorder-average PDF that is itself nearly bimodal, \textit{e.g.}  Fig.~\ref{Histogram}c.
We expect that at even lower temperatures, most PDF should become bimodal, so that the full disorder-averaged distribution then also becomes bimodal.
Bimodal PDF of the overlap have indeed been found in other constrained systems~\cite{B13,KB13,BC14,BJ15,OKIM15}.
We note in passing that a bimodal distribution of overlaps in a cavity is an interesting qualitative signature of the correspondence between the PTS crossover length and a crossover analogous to an equilibrium glass transition in a finite size system.

The shape of the PDF encodes all the information we need about the PTS correlation length.
In particular, the breadth of the distribution can be 
easily detected from the PTS {\it susceptibility}
\be
\chi_{\rm T}\le(R\ri)\equiv \le[\langle \qc^2\rangle_{J(R)}-\langle \qc\rangle_{J(R)}^2\ri]\, .
\ee
The numerical results in Fig.~\ref{PTSs}b
show that this function has a clear non-monotonic evolution with $R$, with a peak that gives an unambiguous, fitting-form-free definition of $\xi_{\rm PTS}$.
In addition, this approach to determining $\xi_{\rm PTS}$ is computationally advantageous because it does not require a careful fit of the PTS correlation tails, where equilibrium sampling is slower within our Monte Carlo scheme, and where the overlap value is small (and hence difficult to precisely measure). 
Using the susceptibility $\chi_{\rm T}$ instead allows us to estimate the PTS length scale from only a handful of very precise measurements around the susceptibility maximum. Given the limited size of the PTS length for most of the computationally-accessible bulk temperatures and the efficiency of parallel-tempering for small cavities, this approach offers an especially effective scheme for determining the PTS length in generic model glass-forming liquids.

It is comforting to note that the peak location of the PTS susceptibility, $\xi_{\rm PTS}^{\rm peak}$, roughly agrees with that obtained through the compressed exponential fit,  $\xi_{\rm PTS}^{\rm fit}$.
The correspondence is also very good with $\xi_{\rm PTS}^{\rm th}$, defined as $G_{\rm PTS}\le(\xi_{\rm PTS}^{\rm th}\ri)\equiv q_{\rm th}$ with $q_{\rm th}=e^{-1}$.
The quantitative agreement between these three definitions 
is demonstrated in Fig.~\ref{PTSs}c, which 
shows that all three estimates give comparable results, except at the (uninteresting) highest temperature studied, $T=1.0$, where the very definition of a PTS length becomes somewhat problematic~\cite{BC14}.
The similarity between the different approaches allows us to provide a reasonable estimate of $\xi_{\rm PTS}$ at the lowest temperature considered, $T=0.45$, which is fairly close to estimates of the mode-coupling crossover temperature for this model, $T_{\rm MCT}\approx0.435$~\cite{KA94}. At that temperature, the susceptibility peak occurs at cavity sizes that are slightly beyond the computational reach of our algorithm. The correlation decay nonetheless provides a reasonable estimate of $\xi_{\rm PTS}$.

Although there are systematic quantitative differences in $\xi_{\rm PTS}$ obtained by the current approach compared with earlier results, these can be explained by the choice of overlap function and other algorithmic details.
Qualitatively, nothing much differs.
The representation of the temperature dependence of $\xi_{\rm PTS}$ in Fig.~\ref{PTSs}c emphasizes the analogy between cavity PTS measurements and other ways of constraining the available phase space of the system that are being actively investigated, such as random pinning~\cite{CB12,JB12,CCT13,CT13,KB13,OKIM15} or coupling to a reference configuration~\cite{FP97,CCGGGPV10,B13,PS14,BJ15,JG15}. 
In all cases, an external constraint with quenched disorder allows one to induce a sharp transition (pinning, coupling) or crossover (cavity) between an unconstrained phase with low overlap and a localized phase with large overlap.
The data in  Fig.~\ref{PTSs}c thus resemble qualitatively published phase diagrams 
for the physics of constrained glass-forming 
liquids~\cite{B13,KB13,BJ15,OKIM15}.
Note that, in all of these approaches, one is interested in understanding the emergence of metastable states in the physically-relevant temperature regime much above the putative Kauzmann transition, whose existence/absence 
is therefore completely immaterial.

\section{Penetration and wandering lengths} 
\label{sec:penperp}

\begin{figure}
\sidesubfloat[]{\includegraphics[width=0.9\textwidth]{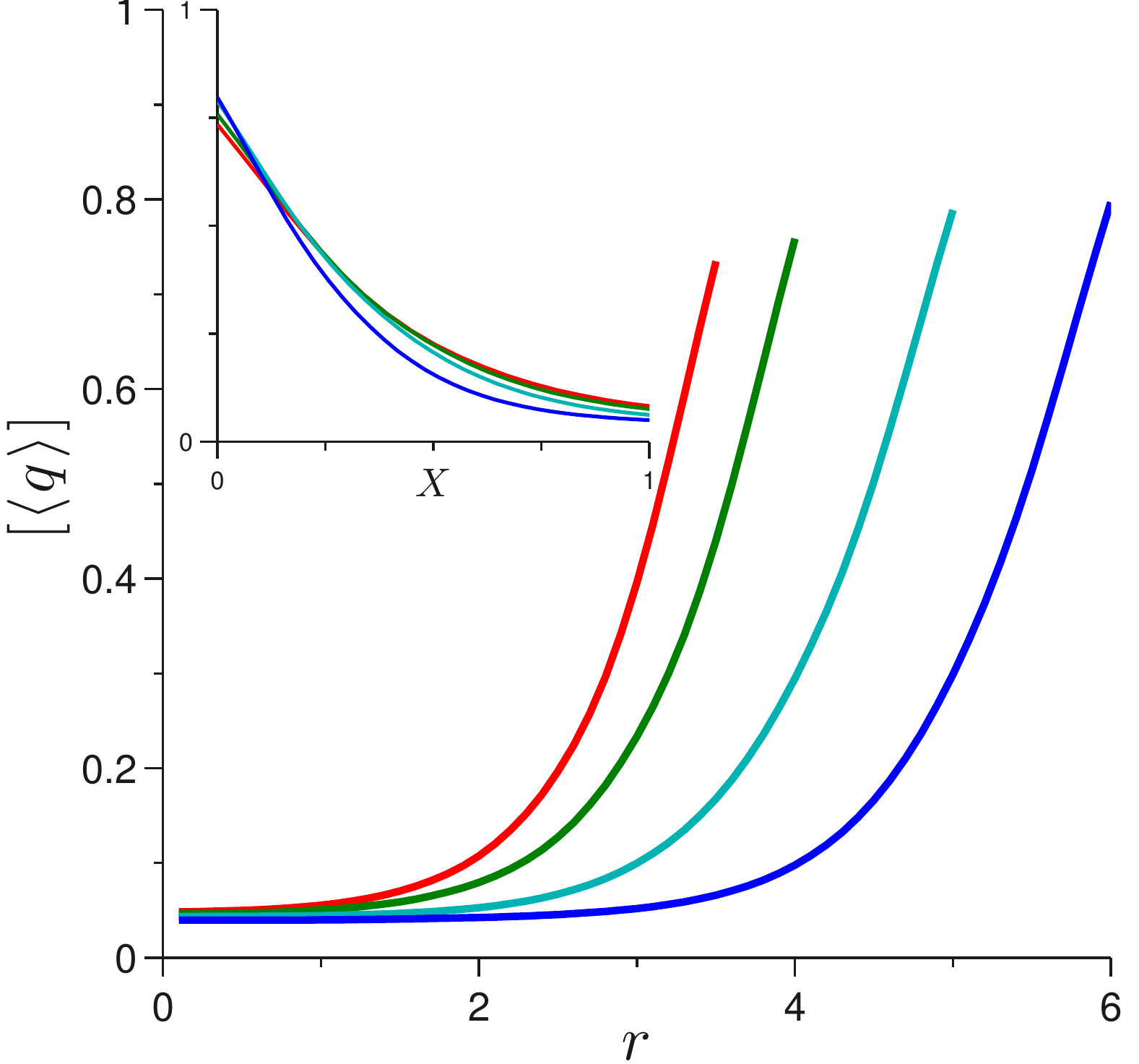}}\quad%
\sidesubfloat[]{\includegraphics[width=0.9\textwidth]{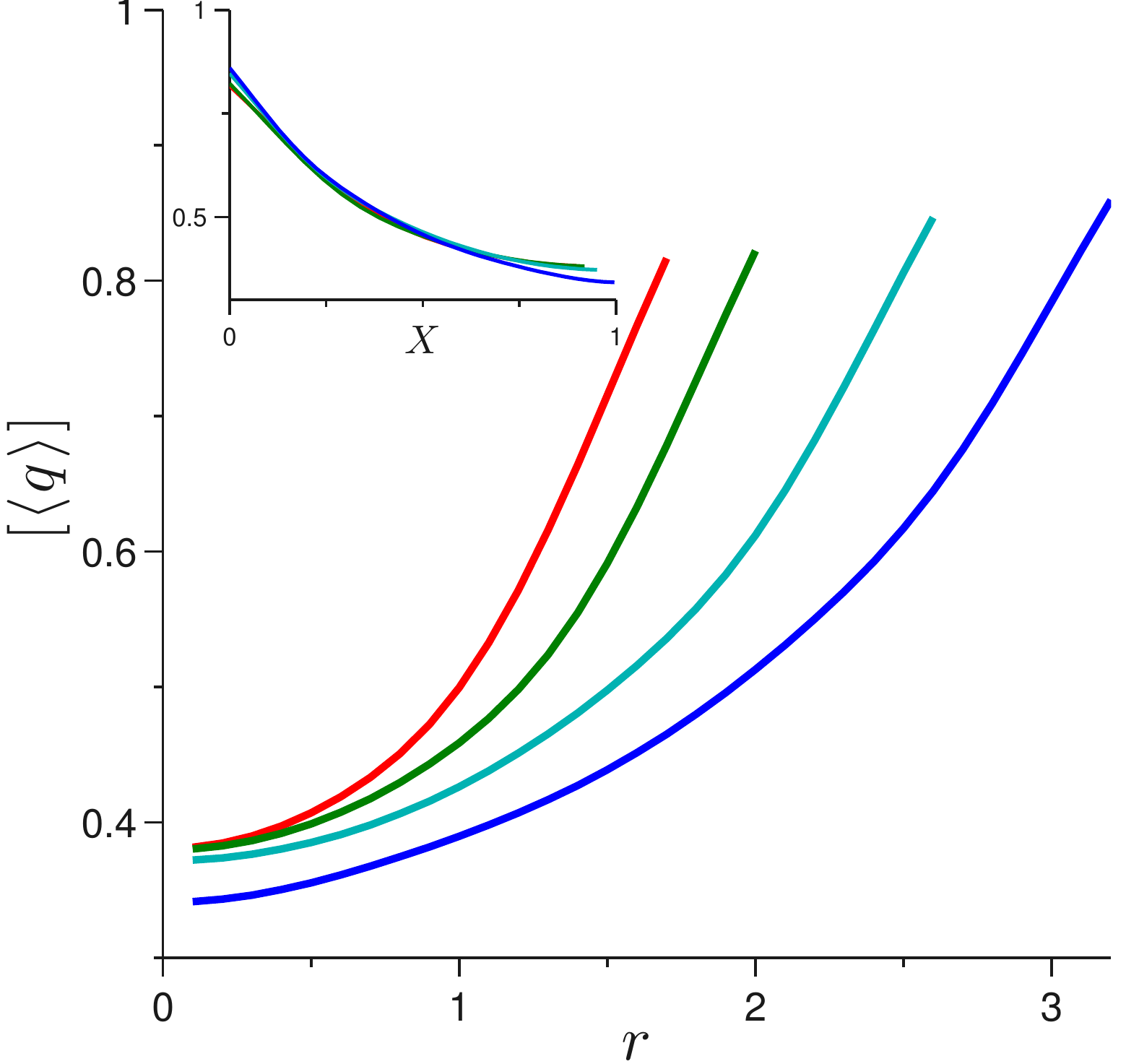}}%
\caption{Spherical  (or $s$-wave) radial profile of $[\langle q\le(\bf{r}\ri) \rangle_{J(R)}]$ for (a) $R\approx\xi_{\rm PTS}$ and (b) $R\approx2\xi_{\rm PTS}$. Colors as in Fig.~\ref{PTSs}. In both panels, insets represent the overlap profile as a function of rescaled variable $X \equiv (R-r)/\xi_{\rm PTS}$, \textit{i.e.}, the distance from the cavity edge rescaled by PTS length. Rescaling nearly collapses the results in (a) but not in (b), which is hinting at a decoupling between $\xi_{\infty}^{\rm wall}$ and $\xi_{\rm PTS}$.}
\label{wall_unbiased}
\end{figure}

The availability of well-averaged configurations for cavities at $R\approx\xi_{\rm PTS}$ enables us to have a closer look into the PTS correlation physics.
We specifically study two length scales associated with the spatial structure of overlap profiles under pinning.
To define them, let us first consider the PTS physics in the limit $R/\xi_{\rm PTS}\rightarrow\infty$, where configurations are frozen beyond a flat wall.
The first length scale of interest is the wall (or penetration) length, $\xi_{\infty}^{\rm wall}$, which corresponds to the characteristic distance from the wall for the decay of the overlap~\cite{CBTT11,KVB12,GTCGV13}.
The spatial fluctuations of this decay length also result in a wandering of the iso-overlap surface.
The second length scale is thus the wandering length, $\xi_{\infty}^{\rm \perp}$, which characterizes the size of these fluctuations in the direction orthogonal to the wall~\cite{BC}.
A priori, there is no relation between the growth of these two scales. The wandering length could even decrease as temperature decreases.
Recent theoretical work~\cite{BC}, however, suggests that these two length scales grow at the same rate, albeit remain subdominant to the cavity PTS correlation length, $\xi_{\rm PTS}$.
Here, we analyze avatars of these two length scales, $\xi_{\star}^{\rm wall}$ and $\xi_{\star}^{\rm \perp}$, in cavities of size $R\approx R_{\star}\equiv\xi_{\rm PTS}$.
We emphasize that these avatars are distinct from their counterparts in the flat wall limit, although they are smoothly connected to $\xi_{\infty}^{\rm wall}$ and $\xi_{\infty}^{\rm \perp}$ as functions of $R/\xi_{\rm PTS}$.

To extract the wall length, we first consider the radial dependence of the overlap profile within a cavity, $[\langle q\le(\bf{r}\ri) \rangle_{J(R)}]$, which can be obtained by binning the overlap profile away from the cavity center (Fig.~\ref{wall_unbiased}a).
When the cavity size $R\approx \xi_{\rm PTS}$, nonlocal constraints imposed by disorder is felt throughout the full cavity, via the rarefaction of metastable states in the local free-energy landscapes.
The near collapse observed in the inset of Fig.~\ref{wall_unbiased}a suggests that the extent of the pinning effect in this regime, $\xi_{\star}^{\rm wall}$, is proportional to $\xi_{\rm PTS}$.
When confined on the scale of $\xi_{\rm PTS}$ the system thus appears fully correlated, as if it were at a sharp transition point.
By contrast, the data in Fig.~\ref{wall_unbiased}b, obtained for $R\approx2\xi_{\rm PTS}$, show that the penetration length measured in a geometry that interpolates between the finite cavity and the flat wall seems to have a milder temperature dependence than $\xi_{\rm PTS}$.
This behavior is as expected, given that $\xi_{\infty}^{\rm wall}$ is observed to grow subdominantly to $\xi_{\rm PTS}$.

\begin{figure}
\centering
\sidesubfloat[]{\includegraphics[width=0.41\textwidth]{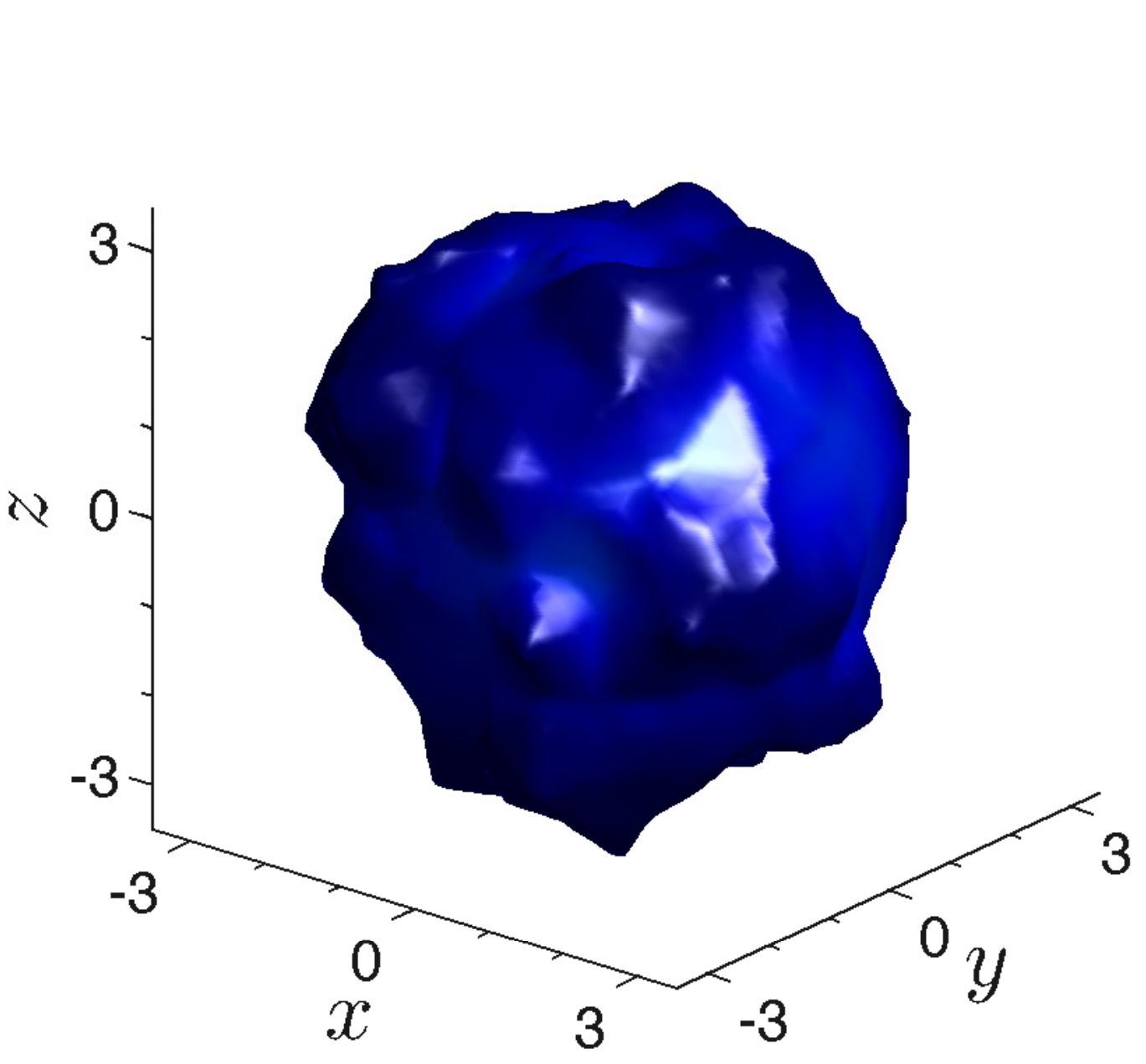}\label{high}}\quad%
\sidesubfloat[]{\includegraphics[width=0.41\textwidth]{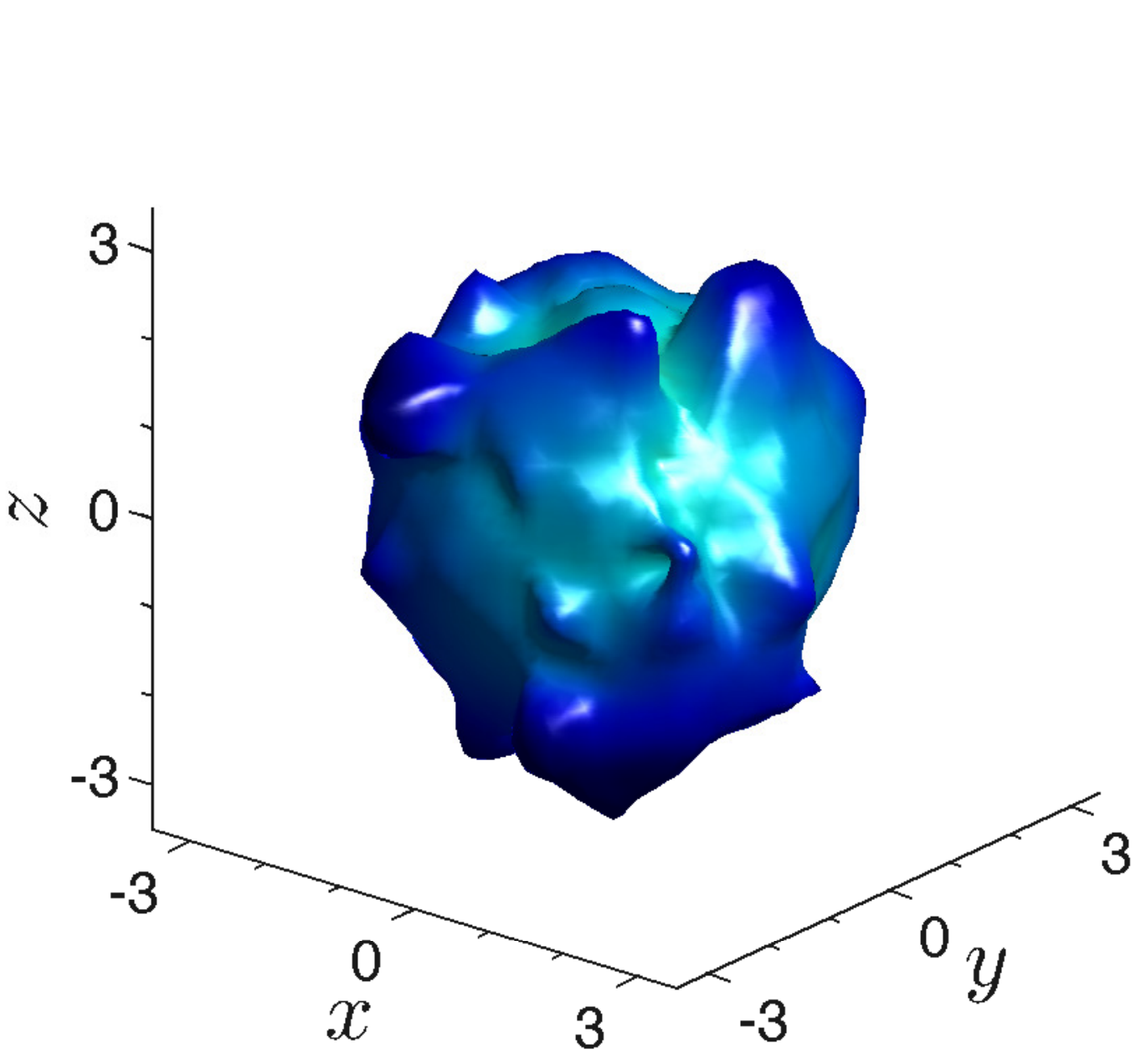}\label{mid}}\quad%
\sidesubfloat[]{\includegraphics[width=0.41\textwidth]{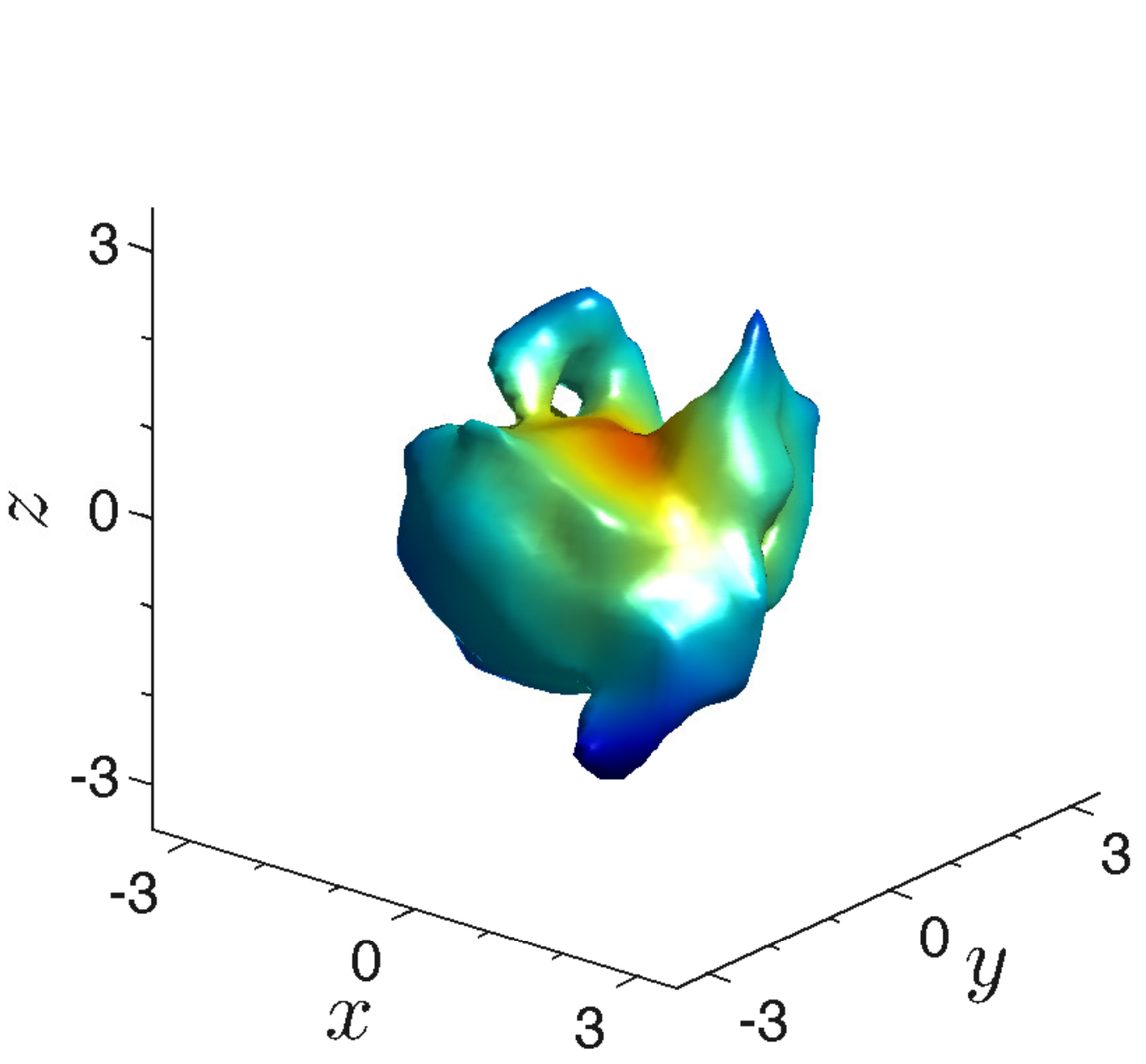}\label{low}}%
\caption{Iso-overlap surfaces $\langle q\le(\bf{r}\ri) \rangle_{J(R)}=q_{\rm iso}$ for $R=3.2$ at $T=0.51$.
Roughness propagates from (a) high overlap $q_{\rm iso}=0.8$ near the cavity edge, to (b) intermediate $q_{\rm iso}=0.6$, and to (c)  low $q_{\rm iso}=0.4$ near the cavity core.}\label{bumpy}
\end{figure}

\begin{figure}
\includegraphics[width=0.9\textwidth]{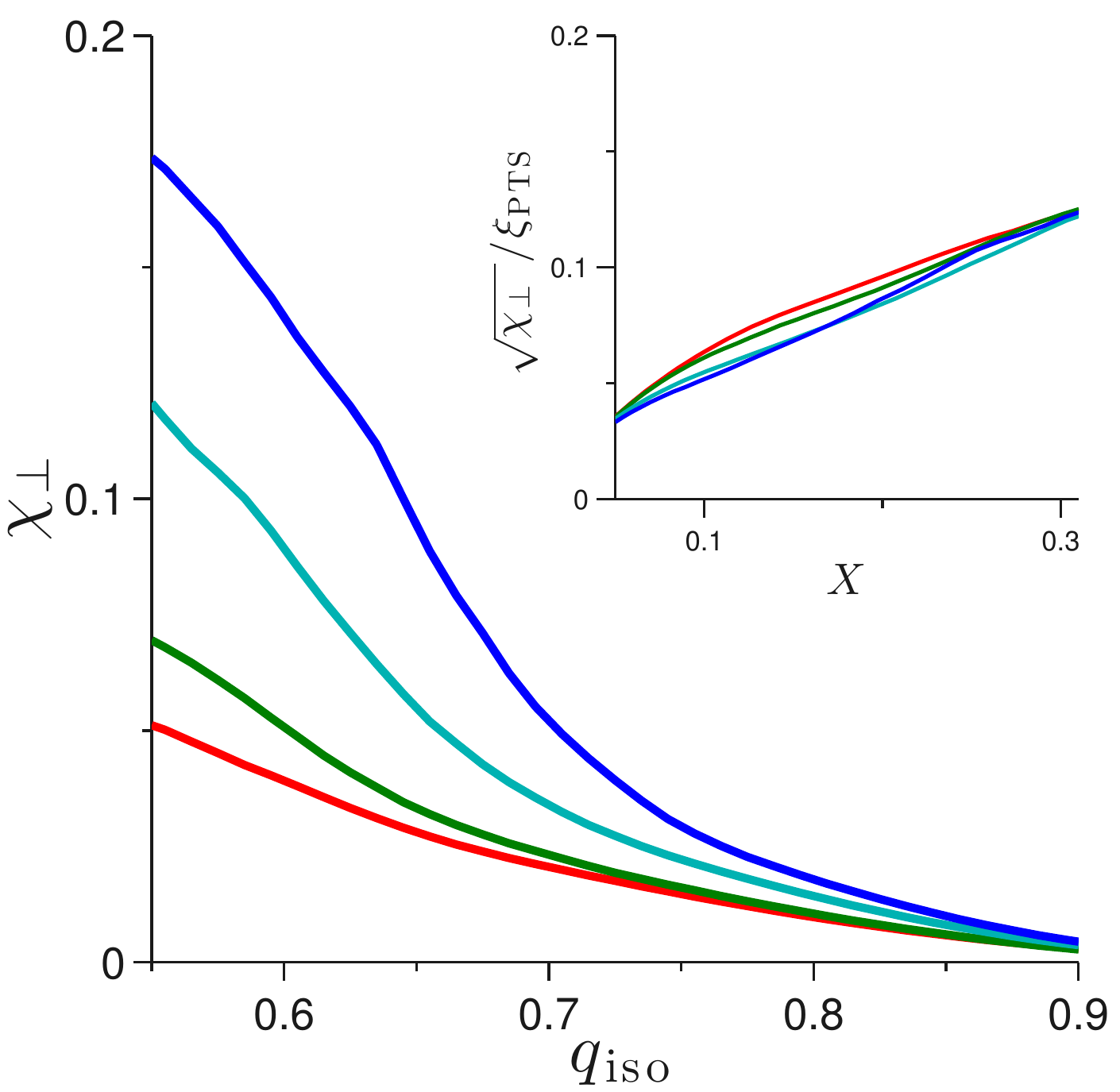}
\caption{Asphericity, $\chi_{\rm \perp}\le(q_{\rm iso}\ri)$, of the iso-overlap surface at various values $q_{\rm iso}$ of the average overlap.
Colors as in Fig.~\ref{PTSs}. For each disorder realization, each increment $\Delta q$ of $q_{\rm iso}$ (see text for details) contains at least $100$ sampling points over the range of $q_{\rm iso}$ shown above.
Inset: Rescaled radial extent of the fluctuating iso-overlap surface $\sqrt{\chi_{\rm \perp}\le(q_{\rm iso}\ri)}$ as a function of rescaled average distance from the cavity edge (see Fig.~\ref{wall_unbiased}) for cavities with $R\approx\xi_{\rm PTS}$. Within the thermal noise, a reasonably good data collapse is observed.}
\label{wander_unbiased}
\end{figure}

To analyze the avatar of wandering length, we consider the orthoradial fluctuations of the overlap field inside the cavity.
Recent field-theoretic calculations  suggest that the overlap profile $\langle q\le(\bf{r}\ri) \rangle_{J(R)}$ for each individual cavity is aspherical~\cite{AALY15,BC}, even though it appears spherically symmetric when averaged over the disorder of the cavity pinning.
A hint of this effect can be seen in Fig.~\ref{bumpy} 
by examining the iso-overlap surfaces
\begin{equation}
\langle q\le(\bf{r}\ri) \rangle_{J(R)}=q_{\rm iso},
\end{equation}
which are generically rather bumpy, even after smoothing particle-scale fluctuations.
In order to quantify asphericity, we calculate $\langle q\le(\bf{r}\ri) \rangle_{J(R)}$ on $10^6$ points uniformly distributed within a cavity. We then consider a point to be part of the iso-overlap surface at $q_{\rm iso}$ if it falls within $[q_{\rm iso}-\frac{\Delta q}{2}, q_{\rm iso}+\frac{\Delta q}{2}]$ with $\Delta q=0.01$, and denote the resulting set of points as $\le\{{\bf r}_i\ri\}_{i=1,...,n}$.
The variance, $\chi_{\rm \perp}\le(q_{\rm iso}\ri)_{J(R)}$, of distances to the edge of the cavity, $\le\{R-\big|{\bf r}_i\big|\ri\}_{i=1,...,n}$, directly quantifies  the sample-to-sample deviations from spherical iso-overlap surfaces. 
The average of this quantity gives an estimate of the distance between the iso-overlap surface and the cavity wall, $R-r\le(q_{\rm iso}\ri)_{J(R)}$. After averaging over the pinning disorder, we obtain
\begin{equation}
\chi_{\rm \perp}\le(q_{\rm iso}\ri)\equiv\le[\chi_{\rm \perp}\le(q_{\rm iso}\ri)_{J(R)}\ri].
\end{equation}
The corresponding numerical results for this quantity at $R\approx\xi_{\rm PTS}$ are shown in Fig.~\ref{wander_unbiased}.
Near the edge of the cavity, $\chi_{\rm \perp}$ is small.
As roughness accumulates toward the core, $\chi_{\rm \perp}$ grows until the iso-overlap surface becomes so rough that, as it is defined, $\chi_{\rm \perp}$ does not approximate $(\xi^{\rm \perp})^2$ well anymore.
Note indeed that in Fig.~\ref{bumpy}c the surface even becomes topologically distinct from a sphere. 
The collapse in the inset of Fig.~\ref{wander_unbiased} suggests again the extent of orthoradial 
fluctuations of the overlap in this regime, $\xi_{\star}^{\rm \perp}$, may be collapsed by the PTS length scale.

On the whole, these results suggest that the spatial structure of overlap profiles in this regime is governed by  a single length scale, $\xi_{\rm PTS}$, although a conclusive interpretation of this effect has yet to be reached.
It would be interesting to examine how this structure morphs into in the flat wall geometry limit, where $\xi_{\infty}^{\rm wall}$ and $\xi_{\infty}^{\rm \perp}$ are expected to grow subdominantly to $\xi_{\rm PTS}$.
The largest cavities we equilibrated in this paper are of the order of $2\xi_{\rm PTS}$, and our preliminary analysis did not conclusively determine whether they belong to the flat wall regime or not, in particular for the wandering length.
A study of the evolution of the overlap profiles for a broader range of cavity sizes, including the careful analysis of the wandering length in the wall geometry, should be performed in future work.

\section{Conclusions}
\label{sec:conclusion}

The results of this work suggest that parallel tempering is a method of choice for studying PTS correlations in generic model glass-forming liquids.
This scheme has indeed allowed us to quantify PTS correlations in a canonical glass former over the temperature regime that is generically studied in the
bulk, bypassing the slowdown due to confinement by amorphous boundaries. 
This scheme thus provides an efficient solution to a major computational obstacle 
that has prevented progress in studies of static correlations in glass formers
for nearly a decade.

In addition, 
by defining a novel overlap susceptibility for finite cavities, we have obtained PTS correlation information well beyond the regime typically attained in simulations, and from a reduced computational effort. Our results are broadly consistent with a growing static cavity PTS length scale as the temperature of a glass-forming liquid is lowered, and thus with glassiness being associated with a roughening of the free energy landscape. 

\begin{figure}
\includegraphics[width=0.9\textwidth]{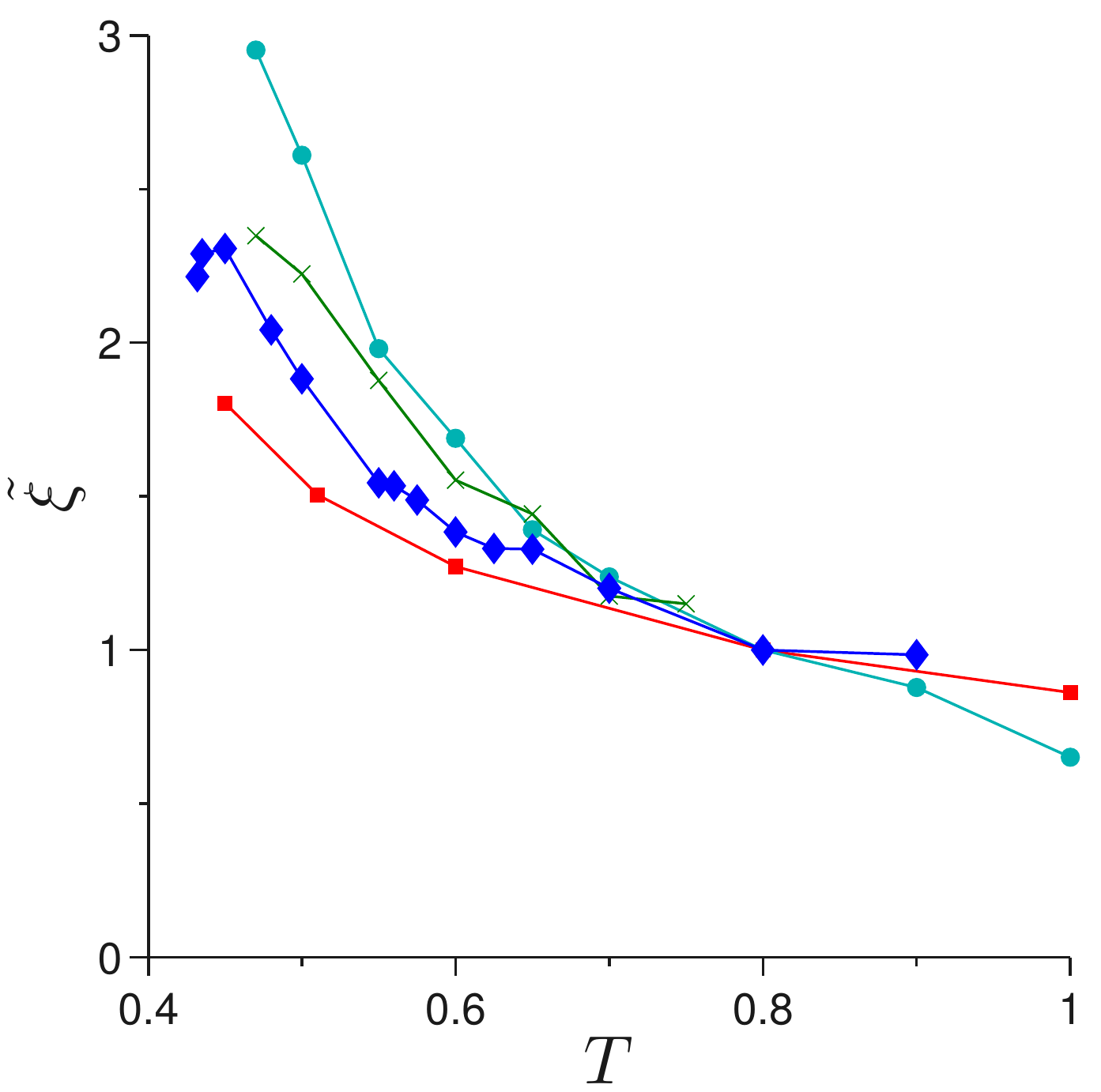}
\caption{Temperature evolution of various normalized length scales, $\tilde{\xi}\equiv\xi/\xi(T_0)$, in the KABLJ model, rescaled to unity at $T_0=0.8$: $\xi_4$ as determined from standard fitting of a four-point dynamic correlation function based on single particle displacements, data taken from Ref.~\onlinecite{FSS14} (cyan-circle) and Ref.~\onlinecite{STK15} (green-cross);
$\xi_{\rm dyn}$ (blue-diamond) obtained from fitting dynamic profiles near an amorphous wall~\cite{HBKR14}; 
and $\xi_{\rm PTS}$ (red-square) is the PTS length scale measured in this work.
A modest decoupling between static and dynamic lengths is observed.}
\label{lengthscale}
\end{figure}

As expected from extrapolating earlier PTS results, however, the 
measured static length scale does not grow by much more than a factor of 
two upon approaching the mode-coupling crossover temperature. 
In Fig.~\ref{lengthscale}, we replot
our data for $\xi_{\rm PTS}$ from Fig.~\ref{PTSs}, and compare its 
temperature evolution with several {\it dynamic} correlation lengthscales.
Given apparent subtleties in extracting dynamical length scales, 
we record three independent 
sets of data taken from Refs.~\onlinecite{FSS14,STK15,HBKR14}, 
taken from either measurements of four-point dynamic 
correlations~\cite{FSS14,STK15} 
or from dynamic profiles near a wall~\cite{HBKR14} for the same model.
This comparison shows that all three dynamic lengths grow more rapidly than the static one toward low temperatures, but the difference is relatively modest.
A stronger decoupling between static and dynamic quantities was observed 
in hard-sphere systems~\cite{BK12,CCT12} than what we find in the KABLJ model. 
The distinction between the two models is in harmony with previous 
findings that signatures of the mode-coupling crossover (associated 
to a rapid growth of a dynamic correaltion length) appear 
weaker in the KABLJ model than in hard or quasi-hard 
spheres~\cite{BBCKT12,HBKR14}.

While there have been recent attempts at experimentally measuring PTS correlations in colloidal glass-formers~\cite{GNGS14,NGSG15}, corresponding studies for molecular glass-formers, roughly $10$ orders of magnitude more sluggish, still remain inaccessible.
Analysis of such extremely sluggish systems would be particularly helpful in properly assessing the theoretical framework that surrounds PTS correlations.
In future work, we will thus couple the present approach with other enhanced sampling techniques in an attempt to push back the computationally accessible boundary for the study of glassiness. 
Such studies will also be useful for systematically investigating the behavior of various length scales $\xi_{\rm PTS}$, $\xi^{\rm wall}_{\star, \infty}$, and $\xi^{\rm \perp}_{\star, \infty}$ and how they relate to one another deep in the dynamically sluggish regime.

\begin{acknowledgments}
We acknowledge many fruitful exchanges and correspondence with 
G.~Biroli, C.~Cammarota, G.~Hocky, T.~Kawasaki, V.~Martin-Mayor, D.~Reichman, G.~Szamel, 
and G.~Tarjus.
We also acknowledge the Duke Compute Cluster for computational resources.
The research leading to these results has received
funding from the European Research Council under
the European Union Seventh Framework Programme
(FP7/2007–2013)/ERC Grant agreement No. 306845 (LB).
PC and SY acknowledge support from the National Science Foundation 
Grant No.~NSF DMR-1055586 and the Sloan Foundation.  
\end{acknowledgments}

\bibliography{glass,HS}

\begin{thebibliography}{48}%
\makeatletter
\providecommand \@ifxundefined [1]{%
 \@ifx{#1\undefined}
}%
\providecommand \@ifnum [1]{%
 \ifnum #1\expandafter \@firstoftwo
 \else \expandafter \@secondoftwo
 \fi
}%
\providecommand \@ifx [1]{%
 \ifx #1\expandafter \@firstoftwo
 \else \expandafter \@secondoftwo
 \fi
}%
\providecommand \natexlab [1]{#1}%
\providecommand \enquote  [1]{``#1''}%
\providecommand \bibnamefont  [1]{#1}%
\providecommand \bibfnamefont [1]{#1}%
\providecommand \citenamefont [1]{#1}%
\providecommand \href@noop [0]{\@secondoftwo}%
\providecommand \href [0]{\begingroup \@sanitize@url \@href}%
\providecommand \@href[1]{\@@startlink{#1}\@@href}%
\providecommand \@@href[1]{\endgroup#1\@@endlink}%
\providecommand \@sanitize@url [0]{\catcode `\\12\catcode `\$12\catcode
  `\&12\catcode `\#12\catcode `\^12\catcode `\_12\catcode `\%12\relax}%
\providecommand \@@startlink[1]{}%
\providecommand \@@endlink[0]{}%
\providecommand \url  [0]{\begingroup\@sanitize@url \@url }%
\providecommand \@url [1]{\endgroup\@href {#1}{\urlprefix }}%
\providecommand \urlprefix  [0]{URL }%
\providecommand \Eprint [0]{\href }%
\providecommand \doibase [0]{http://dx.doi.org/}%
\providecommand \selectlanguage [0]{\@gobble}%
\providecommand \bibinfo  [0]{\@secondoftwo}%
\providecommand \bibfield  [0]{\@secondoftwo}%
\providecommand \translation [1]{[#1]}%
\providecommand \BibitemOpen [0]{}%
\providecommand \bibitemStop [0]{}%
\providecommand \bibitemNoStop [0]{.\EOS\space}%
\providecommand \EOS [0]{\spacefactor3000\relax}%
\providecommand \BibitemShut  [1]{\csname bibitem#1\endcsname}%
\let\auto@bib@innerbib\@empty
\bibitem [{\citenamefont {Berthier}\ and\ \citenamefont {Biroli}(2011)}]{BB11}%
  \BibitemOpen
  \bibfield  {author} {\bibinfo {author} {\bibfnamefont {L.}~\bibnamefont
  {Berthier}}\ and\ \bibinfo {author} {\bibfnamefont {G.}~\bibnamefont
  {Biroli}},\ }\href@noop {} {\bibfield  {journal} {\bibinfo  {journal} {Rev.
  Mod. Phys.}\ }\textbf {\bibinfo {volume} {83}},\ \bibinfo {pages} {587}
  (\bibinfo {year} {2011})}\BibitemShut {NoStop}%
\bibitem [{\citenamefont {Kirkpatrick}, \citenamefont {Thirumalai},\ and\
  \citenamefont {Wolynes}(1989)}]{KTW89}%
  \BibitemOpen
  \bibfield  {author} {\bibinfo {author} {\bibfnamefont {T.~R.}\ \bibnamefont
  {Kirkpatrick}}, \bibinfo {author} {\bibfnamefont {D.}~\bibnamefont
  {Thirumalai}}, \ and\ \bibinfo {author} {\bibfnamefont {P.~G.}\ \bibnamefont
  {Wolynes}},\ }\href@noop {} {\bibfield  {journal} {\bibinfo  {journal} {Phys.
  Rev. A}\ }\textbf {\bibinfo {volume} {40}},\ \bibinfo {pages} {1045}
  (\bibinfo {year} {1989})}\BibitemShut {NoStop}%
\bibitem [{\citenamefont {M\'ezard}\ and\ \citenamefont {Parisi}(1999)}]{MP99}%
  \BibitemOpen
  \bibfield  {author} {\bibinfo {author} {\bibfnamefont {M.}~\bibnamefont
  {M\'ezard}}\ and\ \bibinfo {author} {\bibfnamefont {G.}~\bibnamefont
  {Parisi}},\ }\href@noop {} {\bibfield  {journal} {\bibinfo  {journal} {J.
  Chem. Phys.}\ }\textbf {\bibinfo {volume} {111}},\ \bibinfo {pages} {1076}
  (\bibinfo {year} {1999})}\BibitemShut {NoStop}%
\bibitem [{\citenamefont {Lubchenko}\ and\ \citenamefont
  {Wolynes}(2007)}]{LW07}%
  \BibitemOpen
  \bibfield  {author} {\bibinfo {author} {\bibfnamefont {V.}~\bibnamefont
  {Lubchenko}}\ and\ \bibinfo {author} {\bibfnamefont {P.~G.}\ \bibnamefont
  {Wolynes}},\ }\href@noop {} {\bibfield  {journal} {\bibinfo  {journal} {Ann.
  Rev. Phys. Chem.}\ }\textbf {\bibinfo {volume} {58}},\ \bibinfo {pages} {235}
  (\bibinfo {year} {2007})}\BibitemShut {NoStop}%
\bibitem [{\citenamefont {Bouchaud}\ and\ \citenamefont {Biroli}(2004)}]{BB04}%
  \BibitemOpen
  \bibfield  {author} {\bibinfo {author} {\bibfnamefont {J.-P.}\ \bibnamefont
  {Bouchaud}}\ and\ \bibinfo {author} {\bibfnamefont {G.}~\bibnamefont
  {Biroli}},\ }\href@noop {} {\bibfield  {journal} {\bibinfo  {journal}
  {J.~Chem.~Phys.}\ }\textbf {\bibinfo {volume} {121}},\ \bibinfo {pages}
  {7347} (\bibinfo {year} {2004})}\BibitemShut {NoStop}%
\bibitem [{\citenamefont {Montanari}\ and\ \citenamefont
  {Semerjian}(2006)}]{MS06}%
  \BibitemOpen
  \bibfield  {author} {\bibinfo {author} {\bibfnamefont {A.}~\bibnamefont
  {Montanari}}\ and\ \bibinfo {author} {\bibfnamefont {G.}~\bibnamefont
  {Semerjian}},\ }\href@noop {} {\bibfield  {journal} {\bibinfo  {journal}
  {J.Stat.Phys.}\ }\textbf {\bibinfo {volume} {125}},\ \bibinfo {pages} {23}
  (\bibinfo {year} {2006})}\BibitemShut {NoStop}%
\bibitem [{\citenamefont {Cavagna}, \citenamefont {Grigera},\ and\
  \citenamefont {Verrocchio}(2007)}]{CGV07}%
  \BibitemOpen
  \bibfield  {author} {\bibinfo {author} {\bibfnamefont {A.}~\bibnamefont
  {Cavagna}}, \bibinfo {author} {\bibfnamefont {T.~S.}\ \bibnamefont
  {Grigera}}, \ and\ \bibinfo {author} {\bibfnamefont {P.}~\bibnamefont
  {Verrocchio}},\ }\href@noop {} {\bibfield  {journal} {\bibinfo  {journal}
  {Phys. Rev. Lett.}\ }\textbf {\bibinfo {volume} {98}},\ \bibinfo {eid}
  {187801} (\bibinfo {year} {2007})}\BibitemShut {NoStop}%
\bibitem [{\citenamefont {Biroli}\ \emph {et~al.}(2008)\citenamefont {Biroli},
  \citenamefont {Bouchaud}, \citenamefont {Cavagna}, \citenamefont {Grigera},\
  and\ \citenamefont {Verrocchio}}]{BBCGV08}%
  \BibitemOpen
  \bibfield  {author} {\bibinfo {author} {\bibfnamefont {G.}~\bibnamefont
  {Biroli}}, \bibinfo {author} {\bibfnamefont {J.-P.}\ \bibnamefont
  {Bouchaud}}, \bibinfo {author} {\bibfnamefont {A.}~\bibnamefont {Cavagna}},
  \bibinfo {author} {\bibfnamefont {T.}~\bibnamefont {Grigera}}, \ and\
  \bibinfo {author} {\bibfnamefont {P.}~\bibnamefont {Verrocchio}},\
  }\href@noop {} {\bibfield  {journal} {\bibinfo  {journal} {Nat. Phy.}\
  }\textbf {\bibinfo {volume} {4}},\ \bibinfo {pages} {771} (\bibinfo {year}
  {2008})}\BibitemShut {NoStop}%
\bibitem [{\citenamefont {Sausset}\ and\ \citenamefont {Levine}(2011)}]{SL11}%
  \BibitemOpen
  \bibfield  {author} {\bibinfo {author} {\bibfnamefont {F.}~\bibnamefont
  {Sausset}}\ and\ \bibinfo {author} {\bibfnamefont {D.}~\bibnamefont
  {Levine}},\ }\href@noop {} {\bibfield  {journal} {\bibinfo  {journal} {Phys.
  Rev. Lett.}\ }\textbf {\bibinfo {volume} {107}},\ \bibinfo {pages} {045501}
  (\bibinfo {year} {2011})}\BibitemShut {NoStop}%
\bibitem [{\citenamefont {Hocky}, \citenamefont {Markland},\ and\ \citenamefont
  {Reichman}(2012)}]{HMR12}%
  \BibitemOpen
  \bibfield  {author} {\bibinfo {author} {\bibfnamefont {G.~M.}\ \bibnamefont
  {Hocky}}, \bibinfo {author} {\bibfnamefont {T.~E.}\ \bibnamefont {Markland}},
  \ and\ \bibinfo {author} {\bibfnamefont {D.~R.}\ \bibnamefont {Reichman}},\
  }\href@noop {} {\bibfield  {journal} {\bibinfo  {journal} {Phys. Rev. Lett.}\
  }\textbf {\bibinfo {volume} {108}},\ \bibinfo {pages} {4626} (\bibinfo {year}
  {2012})}\BibitemShut {NoStop}%
\bibitem [{\citenamefont {Biroli}, \citenamefont {Karmakar},\ and\
  \citenamefont {Procaccia}(2013)}]{BKP13}%
  \BibitemOpen
  \bibfield  {author} {\bibinfo {author} {\bibfnamefont {G.}~\bibnamefont
  {Biroli}}, \bibinfo {author} {\bibfnamefont {S.}~\bibnamefont {Karmakar}}, \
  and\ \bibinfo {author} {\bibfnamefont {I.}~\bibnamefont {Procaccia}},\
  }\href@noop {} {\bibfield  {journal} {\bibinfo  {journal} {Phys. Rev. Lett.}\
  }\textbf {\bibinfo {volume} {111}},\ \bibinfo {pages} {165701} (\bibinfo
  {year} {2013})}\BibitemShut {NoStop}%
\bibitem [{\citenamefont {Charbonneau}, \citenamefont {Charbonneau},\ and\
  \citenamefont {Tarjus}(2012)}]{CCT12}%
  \BibitemOpen
  \bibfield  {author} {\bibinfo {author} {\bibfnamefont {B.}~\bibnamefont
  {Charbonneau}}, \bibinfo {author} {\bibfnamefont {P.}~\bibnamefont
  {Charbonneau}}, \ and\ \bibinfo {author} {\bibfnamefont {G.}~\bibnamefont
  {Tarjus}},\ }\href@noop {} {\bibfield  {journal} {\bibinfo  {journal} {Phys.
  Rev. Lett.}\ }\textbf {\bibinfo {volume} {108}},\ \bibinfo {pages} {035701}
  (\bibinfo {year} {2012})}\BibitemShut {NoStop}%
\bibitem [{\citenamefont {Cammarota}\ and\ \citenamefont
  {Biroli}(2012)}]{CB12}%
  \BibitemOpen
  \bibfield  {author} {\bibinfo {author} {\bibfnamefont {C.}~\bibnamefont
  {Cammarota}}\ and\ \bibinfo {author} {\bibfnamefont {G.}~\bibnamefont
  {Biroli}},\ }\href@noop {} {\bibfield  {journal} {\bibinfo  {journal} {Proc.
  Nat. Acad. Sci. U.~S.~A.}\ }\textbf {\bibinfo {volume} {109}},\ \bibinfo
  {pages} {8850} (\bibinfo {year} {2012})}\BibitemShut {NoStop}%
\bibitem [{\citenamefont {Jack}\ and\ \citenamefont {Berthier}(2012)}]{JB12}%
  \BibitemOpen
  \bibfield  {author} {\bibinfo {author} {\bibfnamefont {R.~L.}\ \bibnamefont
  {Jack}}\ and\ \bibinfo {author} {\bibfnamefont {L.}~\bibnamefont
  {Berthier}},\ }\href@noop {} {\bibfield  {journal} {\bibinfo  {journal}
  {Phys. Rev. E}\ }\textbf {\bibinfo {volume} {85}},\ \bibinfo {pages} {021120}
  (\bibinfo {year} {2012})}\BibitemShut {NoStop}%
\bibitem [{\citenamefont {Charbonneau}, \citenamefont {Charbonneau},\ and\
  \citenamefont {Tarjus}(2013)}]{CCT13}%
  \BibitemOpen
  \bibfield  {author} {\bibinfo {author} {\bibfnamefont {B.}~\bibnamefont
  {Charbonneau}}, \bibinfo {author} {\bibfnamefont {P.}~\bibnamefont
  {Charbonneau}}, \ and\ \bibinfo {author} {\bibfnamefont {G.}~\bibnamefont
  {Tarjus}},\ }\href@noop {} {\bibfield  {journal} {\bibinfo  {journal} {J.
  Chem. Phys.}\ }\textbf {\bibinfo {volume} {138}},\ \bibinfo {pages} {12A515}
  (\bibinfo {year} {2013})}\BibitemShut {NoStop}%
\bibitem [{\citenamefont {Charbonneau}\ and\ \citenamefont
  {Tarjus}(2013)}]{CT13}%
  \BibitemOpen
  \bibfield  {author} {\bibinfo {author} {\bibfnamefont {P.}~\bibnamefont
  {Charbonneau}}\ and\ \bibinfo {author} {\bibfnamefont {G.}~\bibnamefont
  {Tarjus}},\ }\href@noop {} {\bibfield  {journal} {\bibinfo  {journal} {Phys.
  Rev. E}\ }\textbf {\bibinfo {volume} {87}},\ \bibinfo {pages} {042305}
  (\bibinfo {year} {2013})}\BibitemShut {NoStop}%
\bibitem [{\citenamefont {Kob}\ and\ \citenamefont {Berthier}(2013)}]{KB13}%
  \BibitemOpen
  \bibfield  {author} {\bibinfo {author} {\bibfnamefont {W.}~\bibnamefont
  {Kob}}\ and\ \bibinfo {author} {\bibfnamefont {L.}~\bibnamefont {Berthier}},\
  }\href@noop {} {\bibfield  {journal} {\bibinfo  {journal} {Phys. Rev. Lett.}\
  }\textbf {\bibinfo {volume} {110}},\ \bibinfo {pages} {245702} (\bibinfo
  {year} {2013})}\BibitemShut {NoStop}%
\bibitem [{\citenamefont {Ozawa}\ \emph {et~al.}(2015)\citenamefont {Ozawa},
  \citenamefont {Kob}, \citenamefont {Ikeda},\ and\ \citenamefont
  {Miyazaki}}]{OKIM15}%
  \BibitemOpen
  \bibfield  {author} {\bibinfo {author} {\bibfnamefont {M.}~\bibnamefont
  {Ozawa}}, \bibinfo {author} {\bibfnamefont {W.}~\bibnamefont {Kob}}, \bibinfo
  {author} {\bibfnamefont {A.}~\bibnamefont {Ikeda}}, \ and\ \bibinfo {author}
  {\bibfnamefont {K.}~\bibnamefont {Miyazaki}},\ }\href@noop {} {\bibfield
  {journal} {\bibinfo  {journal} {Proc. Nat. Acad. Sci. U.S.A.}\ }\textbf
  {\bibinfo {volume} {112}},\ \bibinfo {pages} {6914} (\bibinfo {year}
  {2015})}\BibitemShut {NoStop}%
\bibitem [{\citenamefont {Kob}, \citenamefont {Rold{\'a}n-Vargas},\ and\
  \citenamefont {Berthier}(2012)}]{KVB12}%
  \BibitemOpen
  \bibfield  {author} {\bibinfo {author} {\bibfnamefont {W.}~\bibnamefont
  {Kob}}, \bibinfo {author} {\bibfnamefont {S.}~\bibnamefont
  {Rold{\'a}n-Vargas}}, \ and\ \bibinfo {author} {\bibfnamefont
  {L.}~\bibnamefont {Berthier}},\ }\href@noop {} {\bibfield  {journal}
  {\bibinfo  {journal} {Nat. Phys.}\ }\textbf {\bibinfo {volume} {8}},\
  \bibinfo {pages} {164} (\bibinfo {year} {2012})}\BibitemShut {NoStop}%
\bibitem [{\citenamefont {Hocky}\ \emph {et~al.}(2014)\citenamefont {Hocky},
  \citenamefont {Berthier}, \citenamefont {Kob},\ and\ \citenamefont
  {Reichman}}]{HBKR14}%
  \BibitemOpen
  \bibfield  {author} {\bibinfo {author} {\bibfnamefont {G.~M.}\ \bibnamefont
  {Hocky}}, \bibinfo {author} {\bibfnamefont {L.}~\bibnamefont {Berthier}},
  \bibinfo {author} {\bibfnamefont {W.}~\bibnamefont {Kob}}, \ and\ \bibinfo
  {author} {\bibfnamefont {D.~R.}\ \bibnamefont {Reichman}},\ }\href@noop {}
  {\bibfield  {journal} {\bibinfo  {journal} {Phys. Rev. E}\ }\textbf {\bibinfo
  {volume} {89}},\ \bibinfo {pages} {052311} (\bibinfo {year}
  {2014})}\BibitemShut {NoStop}%
\bibitem [{\citenamefont {Berthier}\ and\ \citenamefont {Kob}(2012)}]{BK12}%
  \BibitemOpen
  \bibfield  {author} {\bibinfo {author} {\bibfnamefont {L.}~\bibnamefont
  {Berthier}}\ and\ \bibinfo {author} {\bibfnamefont {W.}~\bibnamefont {Kob}},\
  }\href@noop {} {\bibfield  {journal} {\bibinfo  {journal} {Phys. Rev. E}\
  }\textbf {\bibinfo {volume} {85}},\ \bibinfo {pages} {011102} (\bibinfo
  {year} {2012})}\BibitemShut {NoStop}%
\bibitem [{\citenamefont {Cavagna}, \citenamefont {Grigera},\ and\
  \citenamefont {Verrocchio}(2012)}]{BICtest12}%
  \BibitemOpen
  \bibfield  {author} {\bibinfo {author} {\bibfnamefont {A.}~\bibnamefont
  {Cavagna}}, \bibinfo {author} {\bibfnamefont {T.~S.}\ \bibnamefont
  {Grigera}}, \ and\ \bibinfo {author} {\bibfnamefont {P.}~\bibnamefont
  {Verrocchio}},\ }\href@noop {} {\bibfield  {journal} {\bibinfo  {journal} {J.
  Chem. Phys.}\ }\textbf {\bibinfo {volume} {136}},\ \bibinfo {pages} {204502}
  (\bibinfo {year} {2012})}\BibitemShut {NoStop}%
\bibitem [{\citenamefont {Cammarota}, \citenamefont {Gradenigo},\ and\
  \citenamefont {Biroli}(2013)}]{CGB13}%
  \BibitemOpen
  \bibfield  {author} {\bibinfo {author} {\bibfnamefont {C.}~\bibnamefont
  {Cammarota}}, \bibinfo {author} {\bibfnamefont {G.}~\bibnamefont
  {Gradenigo}}, \ and\ \bibinfo {author} {\bibfnamefont {G.}~\bibnamefont
  {Biroli}},\ }\href@noop {} {\bibfield  {journal} {\bibinfo  {journal} {Phys.
  Rev. Lett.}\ }\textbf {\bibinfo {volume} {111}},\ \bibinfo {pages} {107801}
  (\bibinfo {year} {2013})}\BibitemShut {NoStop}%
\bibitem [{\citenamefont {Kauzmann}(1948)}]{K48}%
  \BibitemOpen
  \bibfield  {author} {\bibinfo {author} {\bibfnamefont {W.}~\bibnamefont
  {Kauzmann}},\ }\href@noop {} {\bibfield  {journal} {\bibinfo  {journal}
  {Chem. Rev.}\ }\textbf {\bibinfo {volume} {43}},\ \bibinfo {pages} {219}
  (\bibinfo {year} {1948})}\BibitemShut {NoStop}%
\bibitem [{\citenamefont {Berthier}\ and\ \citenamefont
  {Coslovich}(2014)}]{BC14}%
  \BibitemOpen
  \bibfield  {author} {\bibinfo {author} {\bibfnamefont {L.}~\bibnamefont
  {Berthier}}\ and\ \bibinfo {author} {\bibfnamefont {D.}~\bibnamefont
  {Coslovich}},\ }\href@noop {} {\bibfield  {journal} {\bibinfo  {journal}
  {Proc. Nat. Acad. Sci., U.S.A.}\ }\textbf {\bibinfo {volume} {111}},\
  \bibinfo {pages} {11668} (\bibinfo {year} {2014})}\BibitemShut {NoStop}%
\bibitem [{\citenamefont {Kob}\ and\ \citenamefont {Andersen}(1994)}]{KA94}%
  \BibitemOpen
  \bibfield  {author} {\bibinfo {author} {\bibfnamefont {W.}~\bibnamefont
  {Kob}}\ and\ \bibinfo {author} {\bibfnamefont {H.~C.}\ \bibnamefont
  {Andersen}},\ }\href@noop {} {\bibfield  {journal} {\bibinfo  {journal}
  {Phys. Rev. Lett.}\ }\textbf {\bibinfo {volume} {73}},\ \bibinfo {pages}
  {1376} (\bibinfo {year} {1994})}\BibitemShut {NoStop}%
\bibitem [{\citenamefont {Kob}\ and\ \citenamefont {Andersen}(1995)}]{KA95}%
  \BibitemOpen
  \bibfield  {author} {\bibinfo {author} {\bibfnamefont {W.}~\bibnamefont
  {Kob}}\ and\ \bibinfo {author} {\bibfnamefont {H.~C.}\ \bibnamefont
  {Andersen}},\ }\href@noop {} {\bibfield  {journal} {\bibinfo  {journal}
  {Phys. Rev. E}\ }\textbf {\bibinfo {volume} {51}},\ \bibinfo {pages} {4626}
  (\bibinfo {year} {1995})}\BibitemShut {NoStop}%
\bibitem [{\citenamefont {Dyer}, \citenamefont {Lee},\ and\ \citenamefont
  {Yaida}(2013)}]{DLY13}%
  \BibitemOpen
  \bibfield  {author} {\bibinfo {author} {\bibfnamefont {E.}~\bibnamefont
  {Dyer}}, \bibinfo {author} {\bibfnamefont {J.}~\bibnamefont {Lee}}, \ and\
  \bibinfo {author} {\bibfnamefont {S.}~\bibnamefont {Yaida}},\ }\href@noop {}
  {\  (\bibinfo {year} {2013})},\ \Eprint {http://arxiv.org/abs/{\tt
  arXiv:1309.5085}} {{\tt arXiv:1309.5085}} \BibitemShut {NoStop}%
\bibitem [{\citenamefont {Frenkel}\ and\ \citenamefont {Smit}(2001)}]{FS01}%
  \BibitemOpen
  \bibfield  {author} {\bibinfo {author} {\bibfnamefont {D.}~\bibnamefont
  {Frenkel}}\ and\ \bibinfo {author} {\bibfnamefont {B.}~\bibnamefont {Smit}},\
  }\href@noop {} {\emph {\bibinfo {title} {Understanding Molecular
  Simulation}}}\ (\bibinfo  {publisher} {Academic, New York, 2nd Ed.},\
  \bibinfo {year} {2001})\BibitemShut {NoStop}%
\bibitem [{\citenamefont {Fukunishi}, \citenamefont {Watanabe},\ and\
  \citenamefont {Takada}(2002)}]{FWT02}%
  \BibitemOpen
  \bibfield  {author} {\bibinfo {author} {\bibfnamefont {H.}~\bibnamefont
  {Fukunishi}}, \bibinfo {author} {\bibfnamefont {O.}~\bibnamefont {Watanabe}},
  \ and\ \bibinfo {author} {\bibfnamefont {S.}~\bibnamefont {Takada}},\
  }\href@noop {} {\bibfield  {journal} {\bibinfo  {journal} {J. Chem. Phys.}\
  }\textbf {\bibinfo {volume} {116}},\ \bibinfo {pages} {9058} (\bibinfo {year}
  {2002})}\BibitemShut {NoStop}%
\bibitem [{\citenamefont {Fernandez}\ \emph {et~al.}(2009)\citenamefont
  {Fernandez}, \citenamefont {Martin-Mayor}, \citenamefont {Perez-Gaviro},
  \citenamefont {Tarancon},\ and\ \citenamefont {Young}}]{victor1}%
  \BibitemOpen
  \bibfield  {author} {\bibinfo {author} {\bibfnamefont {L.~A.}\ \bibnamefont
  {Fernandez}}, \bibinfo {author} {\bibfnamefont {V.}~\bibnamefont
  {Martin-Mayor}}, \bibinfo {author} {\bibfnamefont {S.}~\bibnamefont
  {Perez-Gaviro}}, \bibinfo {author} {\bibfnamefont {A.}~\bibnamefont
  {Tarancon}}, \ and\ \bibinfo {author} {\bibfnamefont {A.~P.}\ \bibnamefont
  {Young}},\ }\href@noop {} {\bibfield  {journal} {\bibinfo  {journal} {Phys.
  Rev. B}\ }\textbf {\bibinfo {volume} {80}},\ \bibinfo {pages} {024422}
  (\bibinfo {year} {2009})}\BibitemShut {NoStop}%
\bibitem [{\citenamefont {Sciortino}, \citenamefont {Kob},\ and\ \citenamefont
  {Tartaglia}(1999)}]{SKT99}%
  \BibitemOpen
  \bibfield  {author} {\bibinfo {author} {\bibfnamefont {F.}~\bibnamefont
  {Sciortino}}, \bibinfo {author} {\bibfnamefont {W.}~\bibnamefont {Kob}}, \
  and\ \bibinfo {author} {\bibfnamefont {P.}~\bibnamefont {Tartaglia}},\
  }\href@noop {} {\bibfield  {journal} {\bibinfo  {journal} {Phys. Rev. Lett.}\
  }\textbf {\bibinfo {volume} {83}},\ \bibinfo {pages} {3214} (\bibinfo {year}
  {1999})}\BibitemShut {NoStop}%
\bibitem [{\citenamefont {Berthier}(2013)}]{B13}%
  \BibitemOpen
  \bibfield  {author} {\bibinfo {author} {\bibfnamefont {L.}~\bibnamefont
  {Berthier}},\ }\href@noop {} {\bibfield  {journal} {\bibinfo  {journal}
  {Phys. Rev. E}\ }\textbf {\bibinfo {volume} {88}},\ \bibinfo {pages} {022313}
  (\bibinfo {year} {2013})}\BibitemShut {NoStop}%
\bibitem [{\citenamefont {Berthier}\ and\ \citenamefont {Jack}(2015)}]{BJ15}%
  \BibitemOpen
  \bibfield  {author} {\bibinfo {author} {\bibfnamefont {L.}~\bibnamefont
  {Berthier}}\ and\ \bibinfo {author} {\bibfnamefont {R.~L.}\ \bibnamefont
  {Jack}},\ }\href@noop {} {\bibfield  {journal} {\bibinfo  {journal} {Phys.
  Rev. Lett.}\ }\textbf {\bibinfo {volume} {114}},\ \bibinfo {pages} {205701}
  (\bibinfo {year} {2015})}\BibitemShut {NoStop}%
\bibitem [{\citenamefont {Franz}\ and\ \citenamefont {Parisi}(1997)}]{FP97}%
  \BibitemOpen
  \bibfield  {author} {\bibinfo {author} {\bibfnamefont {S.}~\bibnamefont
  {Franz}}\ and\ \bibinfo {author} {\bibfnamefont {G.}~\bibnamefont {Parisi}},\
  }\href@noop {} {\bibfield  {journal} {\bibinfo  {journal} {Phys. Rev. Lett.}\
  }\textbf {\bibinfo {volume} {79}},\ \bibinfo {pages} {2486} (\bibinfo {year}
  {1997})}\BibitemShut {NoStop}%
\bibitem [{\citenamefont {Cammarota}\ \emph {et~al.}(2010)\citenamefont
  {Cammarota}, \citenamefont {Cavagna}, \citenamefont {Giardina}, \citenamefont
  {Gradenigo}, \citenamefont {Grigera}, \citenamefont {Parisi},\ and\
  \citenamefont {Verrocchio}}]{CCGGGPV10}%
  \BibitemOpen
  \bibfield  {author} {\bibinfo {author} {\bibfnamefont {C.}~\bibnamefont
  {Cammarota}}, \bibinfo {author} {\bibfnamefont {A.}~\bibnamefont {Cavagna}},
  \bibinfo {author} {\bibfnamefont {I.}~\bibnamefont {Giardina}}, \bibinfo
  {author} {\bibfnamefont {G.}~\bibnamefont {Gradenigo}}, \bibinfo {author}
  {\bibfnamefont {T.~S.}\ \bibnamefont {Grigera}}, \bibinfo {author}
  {\bibfnamefont {G.}~\bibnamefont {Parisi}}, \ and\ \bibinfo {author}
  {\bibfnamefont {P.}~\bibnamefont {Verrocchio}},\ }\href@noop {} {\bibfield
  {journal} {\bibinfo  {journal} {Phys. Rev. Lett.}\ }\textbf {\bibinfo
  {volume} {105}},\ \bibinfo {pages} {055703} (\bibinfo {year}
  {2010})}\BibitemShut {NoStop}%
\bibitem [{\citenamefont {Parisi}\ and\ \citenamefont {Seoane}(2014)}]{PS14}%
  \BibitemOpen
  \bibfield  {author} {\bibinfo {author} {\bibfnamefont {G.}~\bibnamefont
  {Parisi}}\ and\ \bibinfo {author} {\bibfnamefont {B.}~\bibnamefont
  {Seoane}},\ }\href@noop {} {\bibfield  {journal} {\bibinfo  {journal} {Phys.
  Rev. E}\ }\textbf {\bibinfo {volume} {89}},\ \bibinfo {pages} {022309}
  (\bibinfo {year} {2014})}\BibitemShut {NoStop}%
\bibitem [{\citenamefont {Jack}\ and\ \citenamefont {Garrahan}(2015)}]{JG15}%
  \BibitemOpen
  \bibfield  {author} {\bibinfo {author} {\bibfnamefont {R.~L.}\ \bibnamefont
  {Jack}}\ and\ \bibinfo {author} {\bibfnamefont {J.~P.}\ \bibnamefont
  {Garrahan}},\ }\href@noop {} {\  (\bibinfo {year} {2015})},\ \Eprint
  {http://arxiv.org/abs/{\tt arXiv:1508.06470}} {{\tt arXiv:1508.06470}}
  \BibitemShut {NoStop}%
\bibitem [{\citenamefont {Cammarota}\ \emph {et~al.}(2011)\citenamefont
  {Cammarota}, \citenamefont {Biroli}, \citenamefont {Tarzia},\ and\
  \citenamefont {Tarjus}}]{CBTT11}%
  \BibitemOpen
  \bibfield  {author} {\bibinfo {author} {\bibfnamefont {C.}~\bibnamefont
  {Cammarota}}, \bibinfo {author} {\bibfnamefont {G.}~\bibnamefont {Biroli}},
  \bibinfo {author} {\bibfnamefont {M.}~\bibnamefont {Tarzia}}, \ and\ \bibinfo
  {author} {\bibfnamefont {G.}~\bibnamefont {Tarjus}},\ }\href {\doibase
  10.1103/PhysRevLett.106.115705} {\bibfield  {journal} {\bibinfo  {journal}
  {Phys. Rev. Lett.}\ }\textbf {\bibinfo {volume} {106}},\ \bibinfo {pages}
  {115705} (\bibinfo {year} {2011})}\BibitemShut {NoStop}%
\bibitem [{\citenamefont {Gradenigo}\ \emph {et~al.}(2013)\citenamefont
  {Gradenigo}, \citenamefont {Trozzo}, \citenamefont {Cavagna}, \citenamefont
  {Grigera},\ and\ \citenamefont {Verrocchio}}]{GTCGV13}%
  \BibitemOpen
  \bibfield  {author} {\bibinfo {author} {\bibfnamefont {G.}~\bibnamefont
  {Gradenigo}}, \bibinfo {author} {\bibfnamefont {R.}~\bibnamefont {Trozzo}},
  \bibinfo {author} {\bibfnamefont {A.}~\bibnamefont {Cavagna}}, \bibinfo
  {author} {\bibfnamefont {T.~S.}\ \bibnamefont {Grigera}}, \ and\ \bibinfo
  {author} {\bibfnamefont {P.}~\bibnamefont {Verrocchio}},\ }\href@noop {}
  {\bibfield  {journal} {\bibinfo  {journal} {J. Chem. Phys.}\ }\textbf
  {\bibinfo {volume} {138}},\ \bibinfo {pages} {12A509} (\bibinfo {year}
  {2013})}\BibitemShut {NoStop}%
\bibitem [{\citenamefont {Biroli}\ and\ \citenamefont {Cammarota}(2014)}]{BC}%
  \BibitemOpen
  \bibfield  {author} {\bibinfo {author} {\bibfnamefont {G.}~\bibnamefont
  {Biroli}}\ and\ \bibinfo {author} {\bibfnamefont {C.}~\bibnamefont
  {Cammarota}},\ }\href@noop {} {\  (\bibinfo {year} {2014})},\ \Eprint
  {http://arxiv.org/abs/{\tt arXiv:1411.4566}} {{\tt arXiv:1411.4566}}
  \BibitemShut {NoStop}%
\bibitem [{\citenamefont {Adams}\ \emph {et~al.}(2015)\citenamefont {Adams},
  \citenamefont {Anous}, \citenamefont {Lee},\ and\ \citenamefont
  {Yaida}}]{AALY15}%
  \BibitemOpen
  \bibfield  {author} {\bibinfo {author} {\bibfnamefont {A.}~\bibnamefont
  {Adams}}, \bibinfo {author} {\bibfnamefont {T.}~\bibnamefont {Anous}},
  \bibinfo {author} {\bibfnamefont {J.}~\bibnamefont {Lee}}, \ and\ \bibinfo
  {author} {\bibfnamefont {S.}~\bibnamefont {Yaida}},\ }\href@noop {}
  {\bibfield  {journal} {\bibinfo  {journal} {Phys. Rev. E}\ }\textbf {\bibinfo
  {volume} {91}},\ \bibinfo {pages} {032148} (\bibinfo {year}
  {2015})}\BibitemShut {NoStop}%
\bibitem [{\citenamefont {Flenner}, \citenamefont {Staley},\ and\ \citenamefont
  {Szamel}(2014)}]{FSS14}%
  \BibitemOpen
  \bibfield  {author} {\bibinfo {author} {\bibfnamefont {E.}~\bibnamefont
  {Flenner}}, \bibinfo {author} {\bibfnamefont {H.}~\bibnamefont {Staley}}, \
  and\ \bibinfo {author} {\bibfnamefont {G.}~\bibnamefont {Szamel}},\
  }\href@noop {} {\bibfield  {journal} {\bibinfo  {journal} {Phys. Rev. Lett.}\
  }\textbf {\bibinfo {volume} {112}},\ \bibinfo {pages} {097801} (\bibinfo
  {year} {2014})}\BibitemShut {NoStop}%
\bibitem [{\citenamefont {Shiba}, \citenamefont {Kawasaki},\ and\ \citenamefont
  {Kim}(2015)}]{STK15}%
  \BibitemOpen
  \bibfield  {author} {\bibinfo {author} {\bibfnamefont {H.}~\bibnamefont
  {Shiba}}, \bibinfo {author} {\bibfnamefont {T.}~\bibnamefont {Kawasaki}}, \
  and\ \bibinfo {author} {\bibfnamefont {K.}~\bibnamefont {Kim}},\ }\href@noop
  {} {\  (\bibinfo {year} {2015})},\ \Eprint {http://arxiv.org/abs/{\tt
  arXiv:1510.02546}} {{\tt arXiv:1510.02546}} \BibitemShut {NoStop}%
\bibitem [{\citenamefont {Berthier}\ \emph {et~al.}(2012)\citenamefont
  {Berthier}, \citenamefont {Biroli}, \citenamefont {Coslovich}, \citenamefont
  {Kob},\ and\ \citenamefont {Toninelli}}]{BBCKT12}%
  \BibitemOpen
  \bibfield  {author} {\bibinfo {author} {\bibfnamefont {L.}~\bibnamefont
  {Berthier}}, \bibinfo {author} {\bibfnamefont {G.}~\bibnamefont {Biroli}},
  \bibinfo {author} {\bibfnamefont {D.}~\bibnamefont {Coslovich}}, \bibinfo
  {author} {\bibfnamefont {W.}~\bibnamefont {Kob}}, \ and\ \bibinfo {author}
  {\bibfnamefont {C.}~\bibnamefont {Toninelli}},\ }\href@noop {} {\bibfield
  {journal} {\bibinfo  {journal} {Phys. Rev. E}\ }\textbf {\bibinfo {volume}
  {86}},\ \bibinfo {pages} {031502} (\bibinfo {year} {2012})}\BibitemShut
  {NoStop}%
\bibitem [{\citenamefont {Gokhale}\ \emph {et~al.}(2014)\citenamefont
  {Gokhale}, \citenamefont {Nagamanasa}, \citenamefont {Ganapathy},\ and\
  \citenamefont {Sood}}]{GNGS14}%
  \BibitemOpen
  \bibfield  {author} {\bibinfo {author} {\bibfnamefont {S.}~\bibnamefont
  {Gokhale}}, \bibinfo {author} {\bibfnamefont {K.~H.}\ \bibnamefont
  {Nagamanasa}}, \bibinfo {author} {\bibfnamefont {R.}~\bibnamefont
  {Ganapathy}}, \ and\ \bibinfo {author} {\bibfnamefont {A.~K.}\ \bibnamefont
  {Sood}},\ }\href@noop {} {\bibfield  {journal} {\bibinfo  {journal} {Nat.
  Commun.}\ }\textbf {\bibinfo {volume} {5}},\ \bibinfo {pages} {4685}
  (\bibinfo {year} {2014})}\BibitemShut {NoStop}%
\bibitem [{\citenamefont {Nagamanasa}\ \emph {et~al.}(2015)\citenamefont
  {Nagamanasa}, \citenamefont {Gokhale}, \citenamefont {Sood},\ and\
  \citenamefont {Ganapathy}}]{NGSG15}%
  \BibitemOpen
  \bibfield  {author} {\bibinfo {author} {\bibfnamefont {K.~H.}\ \bibnamefont
  {Nagamanasa}}, \bibinfo {author} {\bibfnamefont {S.}~\bibnamefont {Gokhale}},
  \bibinfo {author} {\bibfnamefont {A.~K.}\ \bibnamefont {Sood}}, \ and\
  \bibinfo {author} {\bibfnamefont {R.}~\bibnamefont {Ganapathy}},\ }\href@noop
  {} {\bibfield  {journal} {\bibinfo  {journal} {Nat. Phys.}\ }\textbf
  {\bibinfo {volume} {11}},\ \bibinfo {pages} {403} (\bibinfo {year}
  {2015})}\BibitemShut {NoStop}%
\bibitem [{\citenamefont {Katzgraber}\ \emph {et~al.}(2006)\citenamefont
  {Katzgraber}, \citenamefont {Trebst}, \citenamefont {Huse},\ and\
  \citenamefont {Troyer}}]{KTHT06}%
  \BibitemOpen
  \bibfield  {author} {\bibinfo {author} {\bibfnamefont {H.~G.}\ \bibnamefont
  {Katzgraber}}, \bibinfo {author} {\bibfnamefont {S.}~\bibnamefont {Trebst}},
  \bibinfo {author} {\bibfnamefont {D.~A.}\ \bibnamefont {Huse}}, \ and\
  \bibinfo {author} {\bibfnamefont {M.}~\bibnamefont {Troyer}},\ }\href@noop {}
  {\bibfield  {journal} {\bibinfo  {journal} {J. Stat. Mech.}\ }\textbf
  {\bibinfo {volume} {2006}},\ \bibinfo {pages} {P03018} (\bibinfo {year}
  {2006})}\BibitemShut {NoStop}%
\end{thebibliography}%

\appendix*

\section{Parallel-tempering parameter choices}

The lists of replica parameters used for each temperature and radius explored in this paper, along with $s_{\rm eq}$, $s_{\rm prod}$, and equilibration success rate are provided in Tables I-V. 
In general, we choose $\le\{(T_a,\lambda_a)\ri\}_{a=1,...,n}$, such that they satisfy (when $n>1$) the linear relation
\be
\frac{T_a-T_1}{T_{\rm dec}-T_1}=\frac{\lambda_a-\lambda_1}{\lambda_{\rm dec}-\lambda_1},
\ee
where $(T_{\rm dec},\lambda_{\rm dec})$ is chosen such that the particles can decorrelate the overlap effectively. We always choose the last replica parameter $\lambda_n\geq\lambda_{\rm dec}$.  

The recording time is fixed as $t_{\rm rec}=10^4$ MC sweeps.
As described in the main text, we discard the first $s_{\rm eq}$ configurations, keep the following $s_{\rm prod}$ configurations, and compare the thermal averages [computed with Eq.~(\ref{qon})]
obtained from two schemes: one starting from the original configuration and the other from a randomized configuration.
The convergence criterion requires that results from both approaches lie within $\pm q_{\rm tol}$ of each other.
For each data point, we record as success rate how many of $50$ cavities satisfy this criterion for a particular $q_{\rm tol}$.
Globally, for $q_{\rm tol}=0.1$, $98\%$ of cavities are deemed to have converged, for all temperatures and radii.

Note that the parameters chosen are neither unique nor optimal.
Some of the chosen parameters result in near bottlenecks in the exchanges of replicas. 
Tuning the replica parameters for each cavity by hand could certainly help sampling low-temperature configurations. A more promising and general way would be to algorithmically tune the replica parameters by monitoring upward and downward flows of replicas~\cite{KTHT06}, in order to reduce the human time investment.

\begin{table}[b]
\begin{tabular}{| l | c | c | c | c | c | c | c | c |}
\hline
\ \ \ $R$ &&\ $1.4$ &\ $1.7$  &\ $2.0$ &\ $2.3$ &\ $2.6$ &\ $2.9$ &\ $3.5$\\
\hline
\hline
success && 100\% & 100\% & 100\% & 100\%  & 100\%  & 100\%    & 100\% \\
\hline
\ \ \ $s_{\rm eq}$ && 1000 & 1000 & 1000 & 1000 & 1000 & 1000 & 1000 \\
\hline
\ \ \ $s_{\rm prod}$ && 4000 & 4000 & 4000 & 4000 & 4000 & 4000 & 4000 \\
\hline
\ \ \ $T_{\rm dec}$ && 1.0000 & 1.0000 & 1.0000 & 1.0000 & 1.0000 & 1.0000 & 1.0000 \\
\hline
\ \ \ $\lambda_{\rm dec}$ && 0.8000 & 0.8000 & 0.8000 & 0.8000 & 0.8000 & 0.8000 & 1.0000 \\
\hline
\ \ $\le\{\lambda_a\ri\}$ && 1.0000 & 1.0000 & 1.0000 & 1.0000 & 1.0000 & 1.0000 & 1.0000 \\
 && 0.9500 & 0.9667 & 0.9750 & 0.9800 & 0.9835 & 0.9870 &  \\
 && 0.9000 & 0.9333 & 0.9508 & 0.9602 & 0.9670 & 0.9741 &  \\
 && 0.8500 & 0.9000 & 0.9271 & 0.9404 & 0.9507 & 0.9610 &  \\
 && 0.8000 & 0.8667 & 0.9025 & 0.9208 & 0.9349 & 0.9484 &  \\
 &&               & 0.8333 & 0.8773 & 0.9015 & 0.9190 & 0.9357 &  \\
 &&               & 0.8000 & 0.8517 & 0.8813 & 0.9025 & 0.9229 &  \\
 &&               &               & 0.8260 & 0.8608 & 0.8862 & 0.9103 &  \\
 &&               &               & 0.7999 & 0.8403 & 0.8696 & 0.8973 &  \\
 &&               &               &              & 0.8200 & 0.8524  & 0.8842 &  \\
 &&               &               &              & 0.7990 & 0.8356  & 0.8707 &  \\
 &&               &               &              &               & 0.8178  & 0.8569 &  \\
 &&               &               &              &               & 0.8000  & 0.8430 &  \\
 &&               &               &              &               &                & 0.8288 &  \\
 &&               &               &              &               &                & 0.8144 &  \\
 &&               &               &              &               &                & 0.8000 &  \\
\hline
\end{tabular}
\caption{Replica parameters for $T=1.00$. Success rate is determined for $q_{\rm th}=0.025$.}
\label{sample2}
\end{table}

\begin{table}
\begin{tabular}{| l | c | c | c | c | c | c | c | c |}
\hline
\ \ \ $R$ &&\ $1.4$ &\ $1.7$  &\ $2.0$ &\ $2.3$ &\ $2.6$ &\ $2.9$ &\ $4.0$\\
\hline
\hline
success && 98\% & 100\% & 98\% & 98\%  & 96\%  & 100\%    & 100\% \\
\hline
\ \ \ $s_{\rm eq}$ && 1000 & 1000 & 1000 & 1000 & 1000 & 1000 & 1000 \\
\hline
\ \ \ $s_{\rm prod}$ && 4000 & 4000 & 4000 & 6000 & 4000 & 4000 & 4000 \\
\hline
\ \ \ $T_{\rm dec}$ && 1.0000 & 1.0000 & 1.0000 & 1.0000 & 0.8000 & 0.8000 & 0.8000 \\
\hline
\ \ \ $\lambda_{\rm dec}$ && 0.8000 & 0.8000 & 0.8000 & 0.8000 & 0.8600 & 0.8800 & 1.0000 \\
\hline
\ \ $\le\{\lambda_a\ri\}$ && 1.0000 & 1.0000 & 1.0000 & 1.0000 & 1.0000 & 1.0000 & 1.0000 \\
 && 0.9550 & 0.9700 & 0.9779 & 0.9812 & 0.9864 & 0.9894 &  \\
 && 0.9100 & 0.9400 & 0.9556 & 0.9625 & 0.9730 & 0.9789 &  \\
 && 0.8650 & 0.9120 & 0.9333 & 0.9444 & 0.9600 & 0.9686 &  \\
 && 0.8200 & 0.8810 & 0.9111 & 0.9265 & 0.9475 & 0.9585 &  \\
 && 0.7750 & 0.8493 & 0.8887 & 0.9086 & 0.9347 & 0.9484 &  \\
 &&               & 0.8165 & 0.8649 & 0.8903 & 0.9208 & 0.9383 &  \\
 &&               & 0.7830 & 0.8409 & 0.8723 & 0.9060 & 0.9283 &  \\
 &&               &               & 0.8167 & 0.8525 & 0.8909 & 0.9174 &  \\
 &&               &               & 0.7923 & 0.8332 & 0.8755 & 0.9054 &  \\
 &&               &               &              & 0.8135 & 0.8600 & 0.8930 &  \\
 &&               &               &              & 0.7935 &               & 0.8800 &  \\
\hline
\end{tabular}
\caption{Replica parameters for $T=0.80$. Success rate is determined for $q_{\rm th}=0.03$.}
\label{sample3}
\end{table}

\begin{table}
\begin{tabular}{| l | c | c | c | c | c | c | c | c | c |}
\hline
\ \ \ $R$ &&\ $1.4$ &\ $1.7$  &\ $2.0$ &\ $2.3$ &\ $2.6$ &\ $2.9$ &\ $3.2$ &\ $5.0$\\
\hline
\hline
success && 100\% & 100\% & 100\% & 98\%  & 100\%  & 98\%   & 98\%  & 100\% \\
\hline
\ \ \ $s_{\rm eq}$ && 1000 & 1000 & 1000 & 1000 & 1000 & 2000 & 2000 & 1000 \\
\hline
\ \ \ $s_{\rm prod}$ && 4000 & 4000 & 4000 & 7000 & 9000 & 8000 & 13000 & 4000 \\
\hline
\ \ \ $T_{\rm dec}$             && 1.0000 & 1.0000 & 1.0000 & 0.8000 & 0.8000 & 0.8000 & 0.8000 & 0.6000 \\
\hline
\ \ \ $\lambda_{\rm dec}$ && 0.8000 & 0.8000 & 0.8000 & 0.8200 & 0.8600 & 0.8800 & 0.8900 & 1.0000 \\
\hline
\ \ $\le\{\lambda_a\ri\}$ && 1.0000 & 1.0000 & 1.0000 & 1.0000 & 1.0000 & 1.0000 & 1.0000 & 1.0000 \\
 && 0.9600 & 0.9725 & 0.9805 & 0.9859 & 0.9877 & 0.9904 & 0.9919 &  \\
 && 0.9200 & 0.9440 & 0.9600 & 0.9721 & 0.9751 & 0.9807 & 0.9839  &  \\
 && 0.8800 & 0.9161 & 0.9400 & 0.9579 & 0.9627 & 0.9711 & 0.9762  & \\
 && 0.8400 & 0.8888 & 0.9201 & 0.9445 & 0.9507 & 0.9619 & 0.9685  & \\
 && 0.7990 & 0.8600 & 0.9000 & 0.9313 & 0.9389 & 0.9528 & 0.9611 &  \\
 &&               & 0.8292 & 0.8790 & 0.9182 & 0.9272 & 0.9435 & 0.9536 &  \\
 &&               & 0.7968 & 0.8567 & 0.9041 & 0.9153 & 0.9340 & 0.9459 &  \\
 &&               &               & 0.8330 & 0.8893 & 0.9022 & 0.9239 & 0.9373 &  \\
 &&               &               & 0.8086 & 0.8731 & 0.8888  & 0.9136 & 0.9283 & \\
 &&               &               & 0.7830 & 0.8560 & 0.8748  & 0.9028 & 0.9191 & \\
 &&               &               &              & 0.8382 & 0.8600  & 0.8915 & 0.9096 &  \\
 &&               &               &              & 0.8200 &                & 0.8800 & 0.8998  & \\
 &&               &               &              &               &                &              & 0.8900 &  \\
\hline
\end{tabular}
\caption{Replica parameters for $T=0.60$. Success rate is determined for $q_{\rm th}=0.06$.}
\label{sample4}
\end{table}

\begin{table}
\begin{tabular}{| l | c | c | c | c | c | c | c | c | c | c |}
\hline
\ \ \ $R$ &&\ $1.4$ &\ $1.7$  &\ $2.0$ &\ $2.3$ &\ $2.6$ &\ $2.9$ &\ $3.2$ &\ $3.5$ &\ $6.0$\\
\hline
\hline
success && 100\% & 100\% & 100\% & 100\%  & 100\%  & 100\%   & 98\%   & 98\%   & 100\% \\
\hline
\ \ \ $s_{\rm eq}$ && 1000 & 1000 & 1000 & 1000 & 1000 & 3000 & 3000 & 4000 & 1000 \\
\hline
\ \ \ $s_{\rm prod}$ && 4000 & 4000 & 4000 & 7000 & 9000 & 7000 & 12000 & 16000 & 4000 \\
\hline
\ \ \ $T_{\rm dec}$             && 1.0000 & 1.0000 & 1.0000 & 0.8000 & 0.8000 & 0.8000 & 0.6000 & 0.6000 & 0.5100 \\
\hline
\ \ \ $\lambda_{\rm dec}$ && 0.8000 & 0.8000 & 0.8000 & 0.8200 & 0.8600 & 0.8800 & 0.8900 & 0.9000 & 1.0000 \\
\hline
\ \ $\le\{\lambda_a\ri\}$ && 1.0000 & 1.0000 & 1.0000 & 1.0000 & 1.0000 & 1.0000 & 1.0000 & 1.0000 & 1.0000 \\
 && 0.9600 & 0.9760 & 0.9825 & 0.9873 & 0.9890 & 0.9916 &  0.9946 &  0.9953 &  \\
 && 0.9200 & 0.9500 & 0.9640 & 0.9744 & 0.9779 & 0.9832 &  0.9892 &  0.9906 &  \\
 && 0.8800 & 0.9240 & 0.9450 & 0.9616 & 0.9668 & 0.9747 &  0.9839 &  0.9861 &  \\
 && 0.8400 & 0.8990 & 0.9250 & 0.9491 & 0.9559 & 0.9666 &  0.9787 &  0.9817 &  \\
 && 0.7970 & 0.8735 & 0.9050 & 0.9372 & 0.9454 & 0.9582 & 0.9737 &  0.9772 &   \\
 &&               & 0.8466 & 0.8850 & 0.9257 & 0.9351 & 0.9500 & 0.9688 &  0.9729 &   \\
 &&               & 0.8178 & 0.8640 & 0.9137 & 0.9243 & 0.9415 & 0.9638 &  0.9688 &   \\
 &&               & 0.7870 & 0.8423 & 0.9011 & 0.9127 & 0.9325 & 0.9590 &  0.9646 &   \\
 &&               &               & 0.8200 & 0.8874 & 0.9005  & 0.9232 & 0.9544 &  0.9605 &   \\
 &&               &               & 0.7960 & 0.8724 & 0.8877  & 0.9132 & 0.9499 &  0.9565 &   \\
 &&               &               &              & 0.8547 & 0.8741  & 0.9026 & 0.9423 &  0.9526 &  \\
 &&               &               &              & 0.8378 & 0.8600  & 0.8915 & 0.9343 &  0.9458 &   \\
 &&               &               &              & 0.8200 &                & 0.8800 & 0.9258 &  0.9386 &  \\
 &&               &               &              &               &                &               & 0.9171 &  0.9311 &  \\
 &&               &               &              &               &                &               & 0.9083 &  0.9235 &  \\
 &&               &               &              &               &                &               & 0.8992 &  0.9159 &  \\
 &&               &               &              &               &                &               & 0.8900 &  0.9080 &  \\
 &&               &               &              &               &                &               &               &  0.9000 &  \\
\hline
\end{tabular}
\caption{Replica parameters for $T=0.51$. Success rate is determined for $q_{\rm th}=0.09$.}
\label{sample6}
\end{table}

\begin{table}
\begin{tabular}{| l | c | c | c | c | c | c | c | c | c |}
\hline
\ \ \ $R$ &&\ $1.4$ &\ $1.7$  &\ $2.0$ &\ $2.3$ &\ $2.6$ &\ $2.9$ &\ $3.2$\\
\hline
\hline
success && 100\% & 100\% & 100\% & 100\%  & 100\%  & 100\%   & 98\% \\
\hline
\ \ \ $s_{\rm eq}$ && 1000 & 1000 & 1000 & 1000 & 2000 & 2000 & 3000 \\
\hline
\ \ \ $s_{\rm prod}$ && 4000 & 4000 & 4000 & 4000 & 6000 & 10000 & 17000 \\
\hline
\ \ \ $T_{\rm dec}$             && 1.0000 & 1.0000 & 1.0000 & 0.8000 & 0.8000 & 0.6000 & 0.6000 \\
\hline
\ \ \ $\lambda_{\rm dec}$ && 0.8000 & 0.8000 & 0.8000 & 0.8200 & 0.8600 & 0.8800 & 0.8900 \\
\hline
\ \ $\le\{\lambda_a\ri\}$ && 1.0000 & 1.0000 & 1.0000 & 1.0000 & 1.0000 & 1.0000 & 1.0000 \\
 && 0.9667 & 0.9760 & 0.9830 & 0.9879 & 0.9913 & 0.9937 & 0.9949 \\
 && 0.9333 & 0.9520 & 0.9650 & 0.9753 & 0.9824 & 0.9876 & 0.9898 \\
 && 0.8950 & 0.9280 & 0.9460 & 0.9627 & 0.9735 & 0.9816 & 0.9847 \\
 && 0.8570 & 0.9040 & 0.9270 & 0.9504 & 0.9645 & 0.9755 & 0.9799 \\
 && 0.8170 & 0.8795 & 0.9070 & 0.9387 & 0.9560 & 0.9693 & 0.9749 \\
 && 0.7710 & 0.8548 & 0.8890 & 0.9273 & 0.9478 & 0.9632 & 0.9699 \\
 &&               & 0.8263 & 0.8680 & 0.9159 & 0.9396 & 0.9574 & 0.9652 \\
 &&               & 0.7960 & 0.8465 & 0.9035 & 0.9312 & 0.9514 & 0.9605 \\
 &&               &               & 0.8235 & 0.8882 & 0.9224  & 0.9458 & 0.9557 \\
 &&               &               & 0.8000 & 0.8729 & 0.9124  & 0.9397 & 0.9506 \\
 &&               &               &              & 0.8560 & 0.8997 & 0.9305 & 0.9427 \\
 &&               &               &              & 0.8381 & 0.8866 & 0.9209 & 0.9345 \\
 &&               &               &              & 0.8200 & 0.8736 & 0.9109 & 0.9259 \\
 &&               &               &              &               & 0.8600 & 0.9006 & 0.9171 \\
 &&               &               &              &               &                & 0.8905 & 0.9084 \\
 &&               &               &              &               &                & 0.8800 & 0.8992 \\
 &&               &               &              &               &                &               & 0.8900 \\
\hline
\end{tabular}
\caption{Replica parameters for $T=0.45$. Success rate is determined for $q_{\rm th}=0.10$.}
\label{sample9}
\end{table}

\end{document}